\documentclass[%
 reprint,
superscriptaddress,
 amsmath,amssymb,
 aps,
 showkeys,
 nolongbibliography,
prb,
]{revtex4-2}
\usepackage{longtable}
\usepackage{adjustbox}

\usepackage{graphicx}
\usepackage{dcolumn}
\usepackage{bm}
\usepackage{float}

\usepackage{lipsum, babel}
\usepackage{footmisc}

\usepackage[table]{xcolor}
\usepackage{tikz}

\usepackage{filecontents}

\begin{document}

\preprint{APS/123-QED}

\title{Reentrant semiconducting behavior in polymerized fullerite structures with increasing sp$^3$-carbon content}

\author{Jorge Laranjeira}
 \email{jorgelaranjeira@ua.pt, lmarques@ua.pt}
 \affiliation{Departamento de Física and CICECO, Universidade de Aveiro, 3810-193 Aveiro, Portugal}
\author{Leonel Marques}%
\email{jorgelaranjeira@ua.pt, lmarques@ua.pt}
\affiliation{Departamento de Física and CICECO, Universidade de Aveiro, 3810-193 Aveiro, Portugal}%

\author{Manuel Melle-Franco}
\affiliation{Departamento de Química and CICECO, Universidade de Aveiro, 3810-193 Aveiro, Portugal}%

\author{Karol Struty\'nski}
\affiliation{Departamento de Química and CICECO, Universidade de Aveiro, 3810-193 Aveiro, Portugal}%

\date{\today}

\begin{abstract}
The electronic behavior of polymerized fullerite structures, ranging from one-dimensional to three-dimensional polymers, was studied using density functional theory. The bandgap across these structures decreases with the rise of sp$^3$-carbon content until metallic behavior is observed. A further increase induces a reopening of the bandgap, revealing a reentrant semiconducting behavior in this class of materials. This behavior is understood in terms of the new electronic states originated by polymeric bonding and the effect of the volume reduction on the dispersion of sp$^2$-states. This study highlights the fullerite polymers as a magnificent platform to tune electronic properties. 		

\end{abstract}


\maketitle

\section*{Introduction}

The electronic bandgap is a key parameter responsible for the optical and electronic properties of materials, making it an important fundamental property to tune in view of possible applications \cite{peplow2013graphene_cit1,bandgap1,bandgap_scuseria}. Bandgap engineering of materials is, therefore, a powerful tool to obtain new physical and chemical properties with possible high technological impact \cite{capasso1987band_cit2,zeghbroeck2007principles_cit3}.

Carbon materials present distinct electronic properties, from semi-metallic behavior in graphite \cite{graphene_old,graphite_tb} to the wide bandgap in diamond \cite{dgap}. Several other carbon allotropes such as peapods, nanotubes and nanoribbons, present a wide range of electronic bandgaps, depending on their structures \cite{encap_enano,notubeselectronic,nanoribons_scuseria,cao2018unconventional,oh2021evidence,kim2022evidence}. Solid C$_{60}$ (fullerite) is a semiconductor and its bandgap decreases under pressure up to 20 GPa, followed by a sudden increase upon further compression \cite{saito1992electric,Vohra_gap,moshary1992gap,regueiro1991gap}. This was rationalized by Saito and coworkers who suggested that the bandgap decrease is the result of the stronger interaction between $\pi$ electrons from neighboring molecules induced by the reduction of the intermolecular distance \cite{saito1992electric}. Conversely, the sudden rise of the bandgap at 20~GPa is ascribed to the molecular collapse with the formation of an amorphous carbon phase having a high content of sp$^3$-hybridized atoms \cite{saito1992electric,Vohra_gap,moshary1992gap}. Similar behavior has also been recently reported in m-xylene solvated C$_{60}$ \cite{c60_xlynes}. 

A family of carbon allotropes that have been less studied regarding the electronic structure are the fullerite polymers, which contain tricoordinated sp$^2$-carbons and tetracoordinated sp$^3$-carbons in a variable ratio. Strong interest in these polymers has resurged quite recently with the synthesis of new frameworks, which could be useful in technological applications, such as the photocatylic water splitting \cite{naturemgc60_23,naturemgc60_22,pan2023long,wang2023}. Most of these phases have been produced recurring to high-pressure high-temperature (HPHT) treatments of non-polymerized fullerite. Low-dimensional polymers are formed at pressures below 8 GPa, in particular one-dimensional (1D) orthorhombic and two-dimensional (2D), tetragonal and rhombohedral, polymers, with all of them containing 66/66 2+2 cycloaddition polymeric bonds and the van der Waals interactions remaining in the non-bonding directions \cite{marques_prl,alvarez}. Above 8~GPa, several three-dimensional (3D) polymerized phases have been synthesized but only few crystalline structures have been proposed so far \cite{alvarez,Laranjeira2017,Laranjeira2019,marquesscience,yamacuboide,yamarombo,BLANK1998}. A face-centered cubic (fcc) 3D polymer was synthesized at 9.5~GPa and 550~$^\circ$C, where the molecules, being in either one of the two standard molecular orientations, are bonded through 56/56 2+2 cycloadditions, the cubic structure resulting from the frustrated arrangement of these bonds in the fcc lattice \cite{Laranjeira2017,Laranjeira2019}. A cuboidal 3D polymerized phase was synthesized by subjecting the 2D tetragonal polymer to 15~GPa and 600 $^\circ$C; its proposed orthorhombic structure involves 6/6 3+3 cycloadditions bonds between neighboring molecules belonging to adjacent (a,b) planes and double 66/66 4+4 cycloadditions along the shortest lattice parameter within the (a,b) plane \cite{yamacuboide}. At the same HPHT conditions, an fcc 3D polymer was synthesized from the non-polymerized fullerite and it was interpreted to be the result of disordering of rhombohedral domains, in which molecules are bonded via 5/5 3+3 cycloadditions to form hexagonal polymerized planes that, in turn, are bonded through 56/65 2+2 cycloadditions \cite{yamarombo}. 

In addition to these fullerite structures, a computationally hypothesized polymerized fullerite clathrate \cite{LARANJEIRA2022} is also to be mentioned since its lattice parameter matches that of an fcc 3D polymer obtained at 12.5 GPa by Brazhkin and coworkers \cite{Braz_clat}. This clathrate structure is constructed by bonding each molecule, adopting a single standard orientation, to the twelve nearest neighbors in the fcc lattice through double 5/5 2+3 cycloaddition bonds. Here, we report systematic electronic structure calculations of the polymerized fullerite structures mentioned above, expanding our previous study \cite{LARANJEIRA2020} and performing more accurate calculations via the Heyd–Scuseria–Ernzerhof (HSE) hybrid functional \cite{bandgap_scuseria,hse}. We show that the bandgap strongly depends on the number of tetracoordinated sp$^3$-carbons exhibited by the polymerized structures until metallic behavior is observed. A further increase leads to a bandgap reopening similar to the reentrant behavior observed in the oxygen-doped carbon nanotubes \cite{reentrant_prl_nanotubos}.

\section*{Methods}
Polymerized fullerite structures were first optimized without any constrain with the Perdew–Burke–Ernzerhof (PBE) \cite{i2,i3} exchange-correlation functional augmented with the classical D3 dispersion term \cite{grimmed3} with Becke-Johnson damping and a 6-31G(d,p) atomic Gaussian basis set, PBE-6-31G(d,p)-D3, as implemented in CRYSTAL17 \cite{crystal17}. To control the Coulomb and exchange infinite lattice series, we utilize five numerical thresholds denoted as T1 to T5. T1 to T4 are set at 12, while T5 is set at 24. The convergence threshold for the self-consistent field (scf) cycle was set to be smaller than $10^{-8}$ Hartree and an energy difference of 10$^{-4}$ Hartree was enforced between consecutive geometric steps. A k-point grid with at least 6×6×6 points was used for the calculations in bulk systems. In addition, when we computed low-dimensional systems, namely 1D chains or 2D self-standing sheets, a minimum of 6 points along each periodic direction was used.

As PBE functional systematically underestimates experimental bandgaps \cite{c60gap,mmarques_hsebetterforsolidswithvasp}, the electronic density of states (DOS) and band structures calculations were computed at the HSE06-6-31G(d,p) level \cite{hse}, which produces improved bandgap values yet at a feasible computational cost \cite{mmarques_hsebetterforsolidswithvasp}. A denser 24×24×24 Monkhorst-Pack \cite{monk_grid} grid and a self-consistent-field convergence threshold of $10^{-10}$ Hartree, were employed for these calculations. Band structure calculations were performed with paths from AFLOW \cite{aflow}.

\section*{Results and Discussion}
\begin{center}
\begin{figure}[H]
	\centering
\includegraphics[scale=0.8]{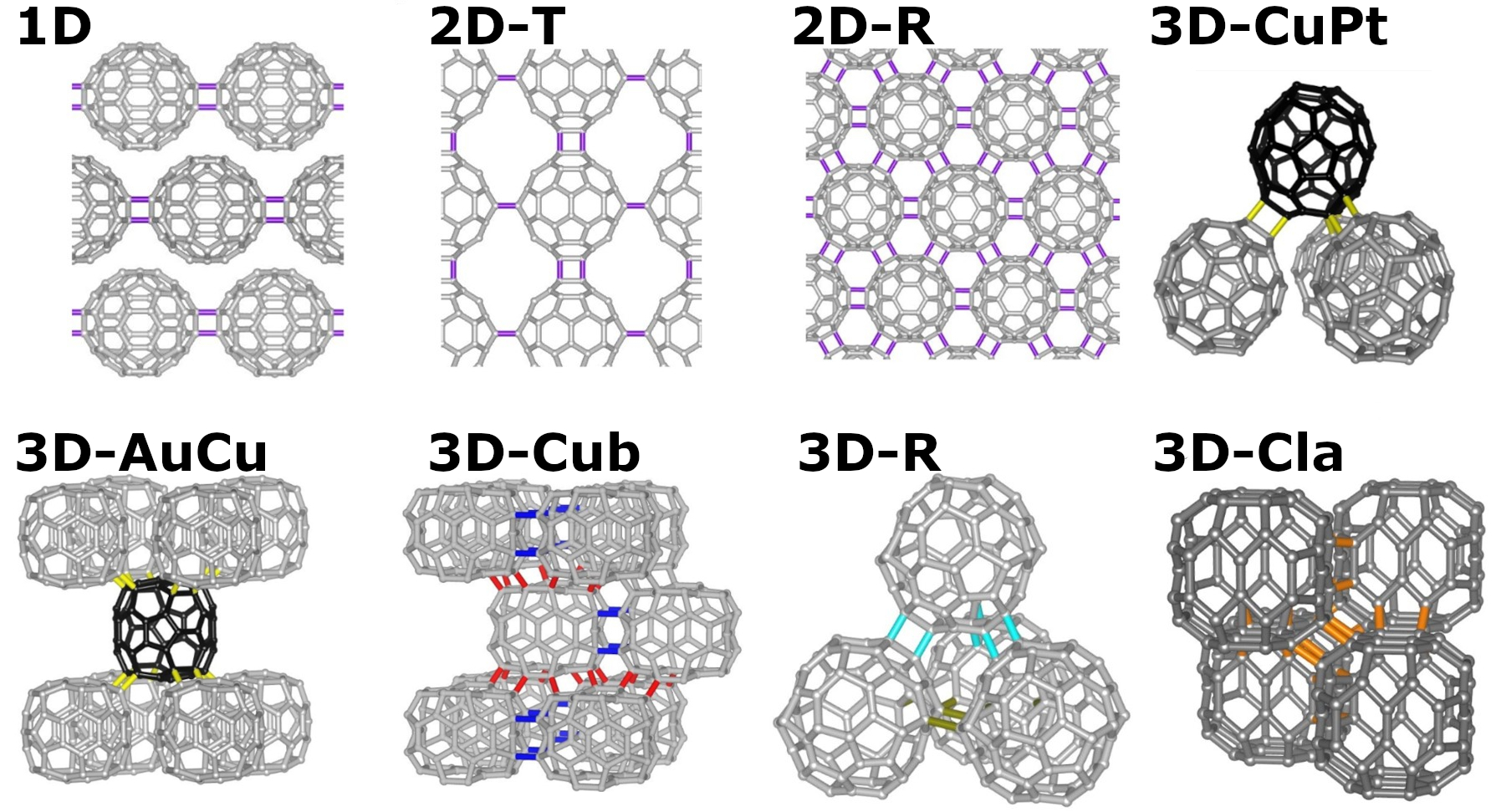}
\caption{Polymerized C$_{60}$ structures. Molecules are drawn in gray and in black for those with different orientations. Different polymeric bonds are drawn in distinct colors: the 66/66 2+2 cycloaddition intermolecular bonds in purple; the 56/56 2+2 cycloaddition in yellow; double 66/66 4+4 cycloadditions are blue-colored; the 6/6 3+3 cycloaddition bonds in red, 5/5 3+3 cycloaddition are green-colored bonds; the 56/65 2+2 cycloaddition bonds in light blue and the orange-colored bonds are double 5/5 2+3 cycloadditions.}
	\label{struc}
\end{figure}
\end{center}

All the studied crystal structures are depicted in figure \ref{struc} and their Wyckoff positions are given in the Supporting Information (SI), section A. The electronic band structures and densities of states (DOS) are also shown in SI, section B. Table \ref{at3} gives a summary of the optimized structural properties and electronic bandgaps of the polymerized fullerite structures considered in this study. The van der Waals fullerite (non-polymerized fullerite), at room pressure and temperature, is taken as a reference and for simplicity is denoted as 0D \cite{fleming}. The 3D-AuCu-type and 3D-CuPt-type structures are ordered configurations of the frustrated fcc 3D polymer synthesized at 9.5 GPa and are described in detail elsewhere \cite{laran2018}.

 \onecolumngrid
\twocolumngrid

\onecolumngrid
\LTcapwidth=\textwidth
\begin{center}
\begin{longtable}{c | c | c | c | c | c} 
  \caption{Optimized lattice parameters and volume per molecule for the polymerized fullerite structures, at the PBE-6-31G(d,p)-D3 level. The space group, the number of tetracoordinated sp$^3$-carbons per molecule and the electronic bandgap, calculated at the HSE-6-31G(d,p) theory level, are also presented for each structure. The 3R hexagonal lattice is used to describe the rhombohedral structures.}\\ 
	\label{at3}\\
structure & space group & DFT cell constants (\AA)  & V(\AA$^3$/C$_{60}$) & \#sp$^3$-atoms/C$_{60}$ & Bandgap (eV)  \\
			         
			\hline 
                                &             &  a=14.08                     &       &         &                 \\
     0D                         & Fm$\bar{3}$ &  b=14.08                     & 698.26&     0   & 1.5407          \\
                                &             &  c=14.08                     &       &         &                 \\
                                &             & $\alpha=\beta=\gamma=90^\circ$     &       &         &                 \\
    \hline                                                                                   
                                &             & a=9.11                       &       &         &                 \\
     1D                         &  Pmnn       & b=9.92                       & 656.69&     4   & 1.3326          \\
                                &             & c=14.54                      &       &         &                 \\
                                &             & $\alpha=\beta=\gamma=90^\circ$     &       &         &                 \\
    \hline                                                                             
                                &             & a=9.05                       &       &         &                 \\
    2D-tetragonal               &  Immm       & b=9.15                       & 617.96&     8   & 1.3771          \\
                                &             & c=14.91                      &       &         &                 \\
                                &             & $\alpha=\beta=\gamma=90^\circ$     &       &         &                 \\
    \hline                                                                            
                                &             &  a=9.19                      &       &         &                 \\
    2D-rhombohedral             & R$\bar{3}$m &  b=9.19                      & 601.45&   12    & 1.1479          \\
                                &             &  c=24.65                     &       &         &                 \\
                                &             & $\alpha=\beta=90^\circ;\gamma=120^\circ$ &       &         &                 \\
    \hline                                                                            
                                &             &   a=9.45                     &       &         &                 \\
    3D-CuPt-type                & R$\bar{3}$c &   b=9.45                     & 576.22&   12    & 0.0628          \\
                                &             &   c=2x22.34                  &       &         &                 \\
                                &             & $\alpha=\beta=90^\circ;\gamma=120^\circ$ &       &         &                 \\
    \hline                                                                        
                                &             & a=9.27                       &       &         &                 \\
    3D-AuCu-type                & P4$_2$/mnm  & b=9.27                       & 553.10&   16    & --              \\
                                &             & c=12.87                      &       &         &                 \\
                                &             & $\alpha=\beta=\gamma=90^\circ$     &       &         &                 \\
    \hline                                                                            
                                &             &  a=8.50                      &       &         &                 \\
     3D-cuboidal                & Immm        &  b=8.62                      & 481.92&   24    & --              \\
                                &             &  c=13.16                     &       &         &                 \\
                                &             & $\alpha=\beta=\gamma=90^\circ$     &       &         &                 \\
    \hline                                                                            
                                &             &                       &       &         &                 \\
                                &             &   a=8.92                     &       &         &                 \\
    3D-rhombohedral             & R$\bar{3}$ &   b=8.92                      & 513.48&   24    & 1.3796          \\
                                &             &   c=22.35                    &       &         &                 \\
                                &             & $\alpha=\beta=90^\circ;\gamma=120^\circ$ &       &         &                 \\
    \hline                                                                             
                                &             &   a=12.41                    &       &         &                 \\
    3D-clathrate                & Pm$\bar{3}$ &   b=12.41                    & 478.31&   48    & 1.2806          \\
                                &             &   c=12.41                    &       &         &                 \\
                                &             & $\alpha=\beta=\gamma=90^\circ$     &       &         &                 \\
                    
                                      \hline
	\end{longtable}
 \end{center}
\twocolumngrid

\begin{center}
\begin{figure}[H]
		\hspace*{-0.5cm} 
	\centering
\includegraphics[scale=0.7]{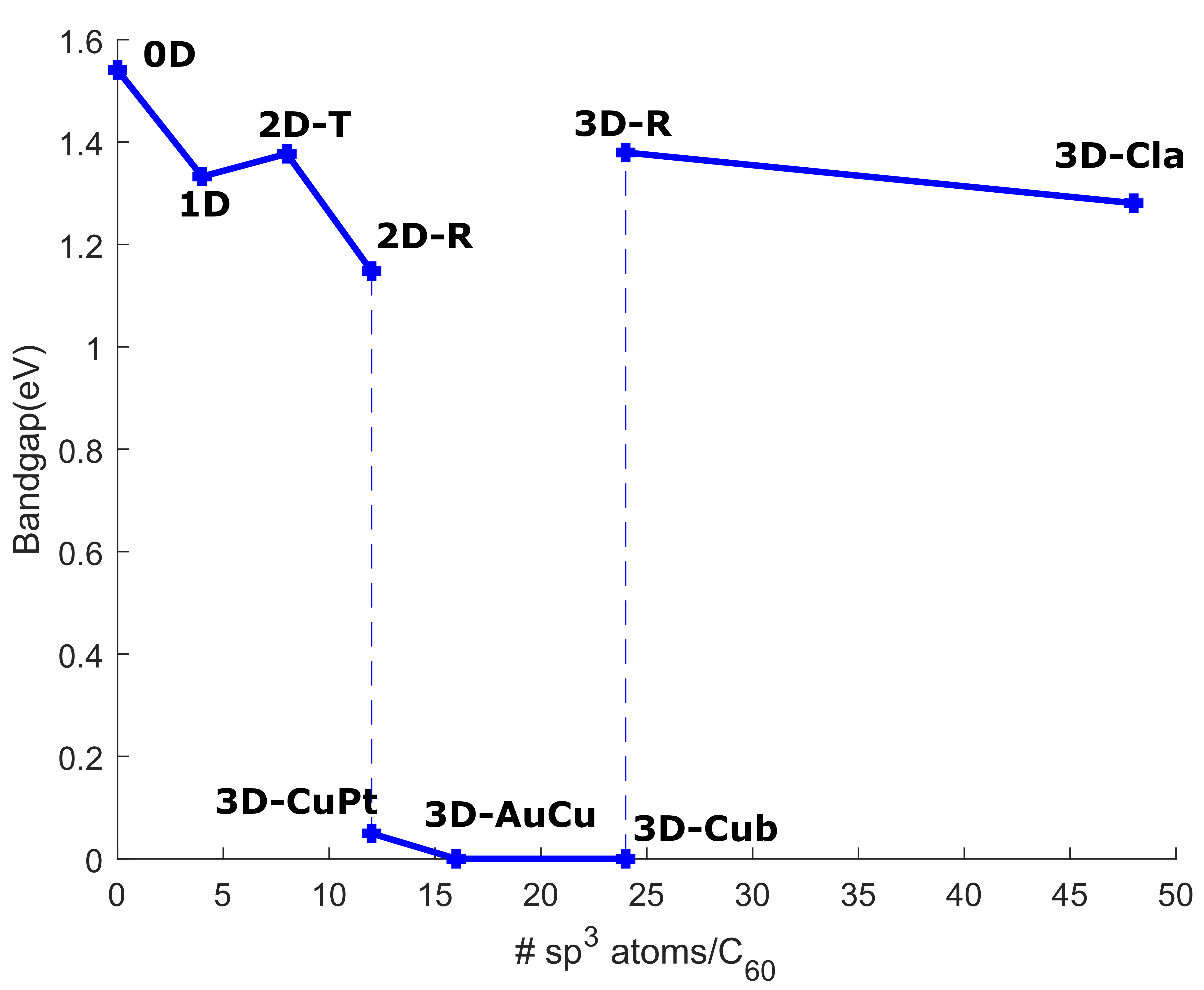}
	\caption{Electronic bandgap against the number of sp$^3$-carbons on each molecule for the bulk polymerized structures. The blue line serves as a guide to the eye.}
	\label{gaps_evol}
\end{figure}
\end{center}

The overall electronic behavior of the polymerized structures is shown in figure \ref{gaps_evol}, where their bandgaps are plotted against the number of tetracoordinated sp$^3$-carbons on each molecule, i.e. the number of polymeric bonds. Three zones can be easily distinguished: initially the bandgap shows a gradual decrease, for the low-dimensional 1D and 2D polymerized structures, followed by its closure (metallicity), for the 3D polymerized structures with moderate sp$^3$-carbon content, and finally the bandgap resurges for the 3D polymerized structures with high sp$^3$-carbon content. This plot illustrates quite well a marked reentrant semiconducting behavior in this class of materials. 

In the first zone, as referred, the electronic bandgap essentially decreases with the number of sp$^3$-carbons exhibited by the low-dimensional polymerized structures. However, this tendency is not followed when going from the 1D structure to the 2D tetragonal structure, which has a higher bandgap while possessing a higher number of sp$^3$-carbons. In order to gain insight about this deviant behavior and about the influence of the van der Waals description on the electronic properties of the low-dimensional polymers, the bandgaps were computed for the corresponding self-standing systems, namely a molecule, a chain, a quadratic polymerized layer and an hexagonal polymerized layer, and are plotted in the SI figure S10. These systems show larger bandgaps than the corresponding bulk structures but, in contrast, the bandgap decreases monotonously with the number of sp$^3$-carbons; for instance, the polymerized chain has a bandgap of 1.85 eV, larger than that of the quadratic polymerized layer, 1.55 eV. Further confirmation of this general tendency is obtained when the bandgaps are computed from the structures that were optimized keeping the lattice parameters constrained to the experimental values (see figure S11) instead of using the full-optimized, and more compacted, structures.

\begin{center}
\begin{figure}[H]
	\centering
         \includegraphics[scale=0.8]{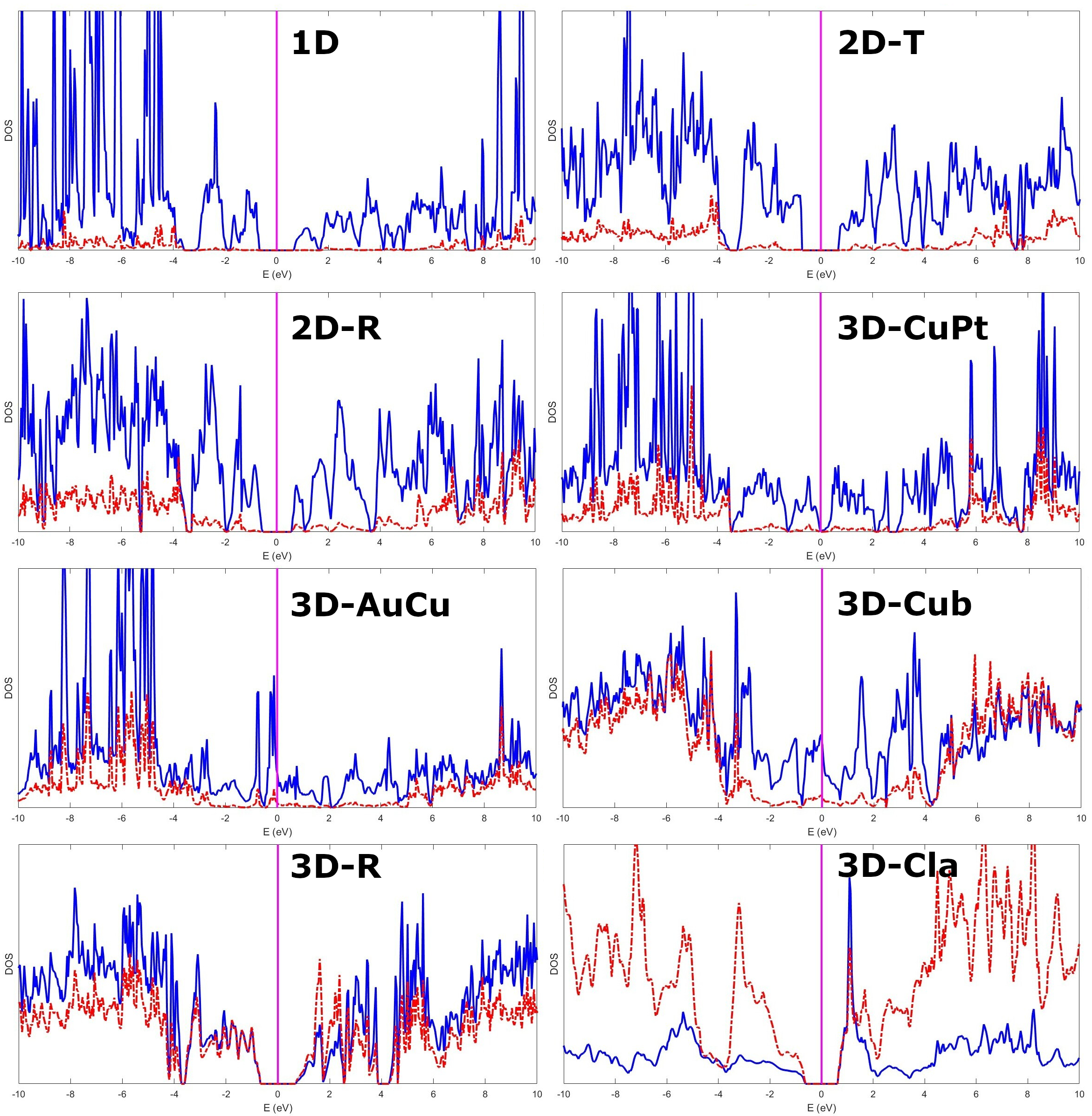}
	\caption{The electronic density of states projected (PDOS) in the sp$^2$- and sp$^3$-hybridized carbons for the polymerized structures. The blue line corresponds to sp$^2$-carbons while the red-dotted line corresponds to sp$^3$-carbons. The vertical magenta line indicates the Fermi level.}
	\label{pardos}
\end{figure} 
\end{center}

\begin{figure}[ht!]
	\hspace*{-0.5cm} 
 	\includegraphics[scale=0.7]{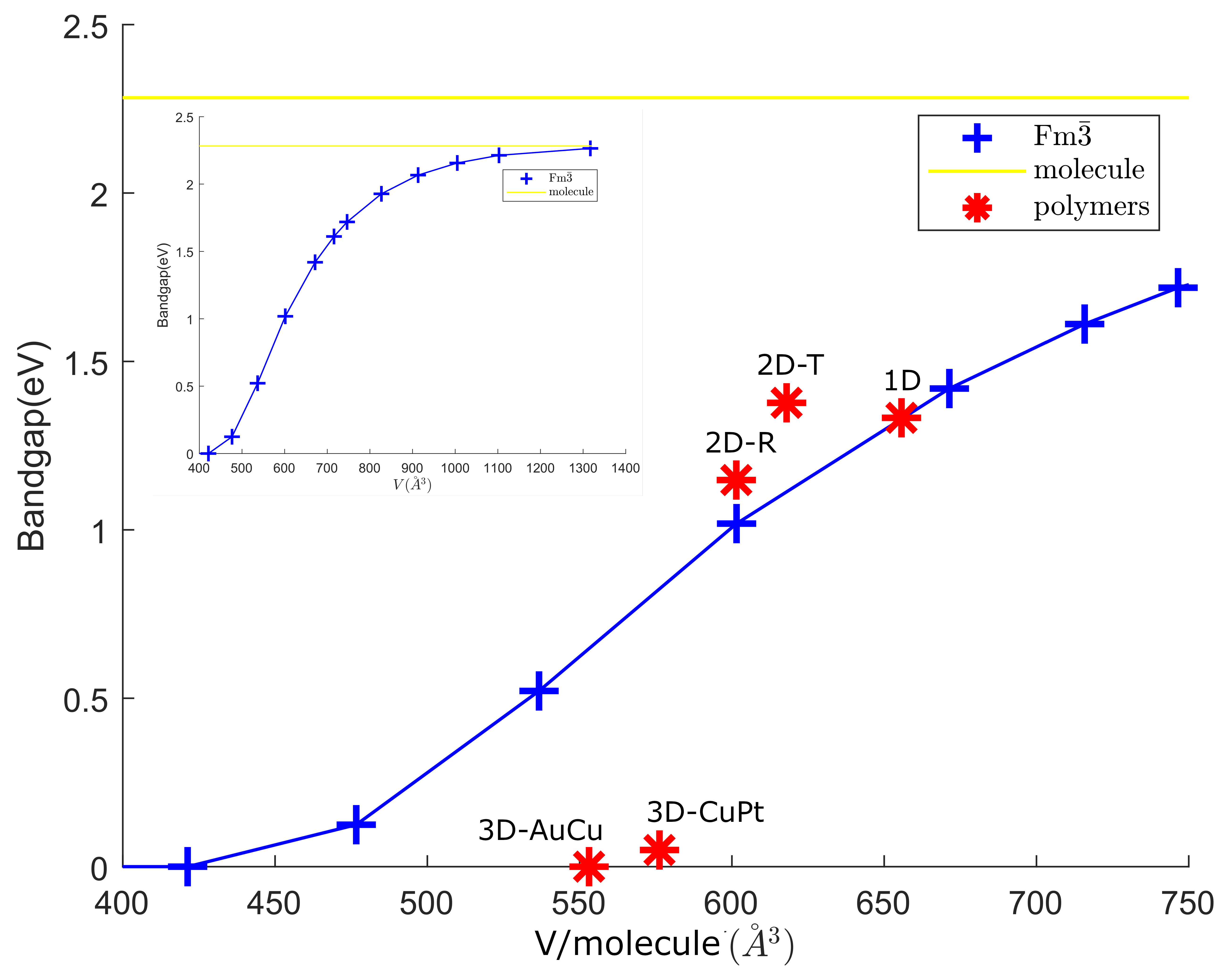}
 	\caption{The bandgap of the non-polymerized fullerite structure, with space group Fm$\bar{3}$, calculated at different volumes up to the bandgap closure. The blue line serves as a guide to the eye. Some polymerized structures are also added to the plot for comparison. The yellow line indicates the bandgap for the self-standing molecule. Inset: same curve for a larger volume domain.}
	\label{sp3carbonds_dim}
\end{figure}

The sp$^3$/sp$^2$ projected density of states (PDOS) curves, given in figure \ref{pardos}, indicate that the sp$^3$-carbons arising from the polymeric bonding barely contribute to the valence and conduction bands of the low-dimensional polymers. Additional support to this conclusion comes from figure \ref{sp3carbonds_dim}, where the almost-linear decline of the calculated bandgap with decreasing volume, observed in the non-polymerized fullerite (in which tetracoordinated sp$^3$-carbons are absent), is nearly followed by the low-dimensional polymerized structures. This behavior, experimentally observed a long time ago in non-polymerized fullerite, is due to the stronger interaction of the $\pi$ molecular orbitals under compression that causes an increased dispersion of the valence and conduction bands and, consequently, a bandgap decrease \cite{saito1992electric}. This interpretation is, then, likely to also apply to the low-dimensional polymerized structures, the decrease of the bandgap being primarily a consequence of the volume reduction (induced by the formation of polymeric bonds) on pure sp$^2$-states and not due to new electronic states arising from the sp$^3$-carbons. 


The binary-alloy type structures, 3D-CuPt-type and 3D-AuCu-type, in which the polymeric bonding extends in three dimensions, display almost metallic and metallic behavior, respectively. In opposite to the low-dimensional polymers, the contribution of the sp$^3$-carbons to electronic states is now present well into the Fermi region, as can be explicitly seen in the PDOS curves of figure \ref{pardos}, being this contribution crucial to close the bandgap and occurrence of metallicity. However, as will be discussed below, the appearance of metallicity occurs only when
the states in the Fermi region are not dominated by the contribution of sp$^3$-carbons. The importance of the sp$^3$-contributed states to the occurrence of metallicity is further emphasized by the results of figure \ref{sp3carbonds_dim}, which shows that the bandgap of the non-polymerized fullerite (with no tetracoordinated sp$^3$-carbons) closes at a volume much smaller than those of these metallic polymerized structures. Thus, in contrast to low-dimensional polymers, the electronic behavior of these 3D polymers, in particular the observation of metallicity, cannot be simply a consequence of the volume reduction on pure sp$^2$-states. The first experimental observation of metallic behavior in C$_{60}$ polymers was reported by Buga and coworkers, for a 3D phase prepared at 9.5 GPa and 550 $^\circ$C \cite{BUGA_1,BUGA_2}.

Another 3D polymerized fullerite showing metallic behavior is the so-called cuboidal structure \cite{yamacuboide}. In contrast, the 3D-rhombohedral polymerized structure is semiconducting, despite having the same number of sp$^3$-carbons \cite{yamarombo}. The PDOS curves given in figure \ref{pardos}, provide insight on these different electronic properties. For the 3D-rhombohedral structure, the states with a larger contribution of sp$^3$-carbons are mostly concentrated around the Fermi level determining its semiconducting behavior, and the states with a larger contribution of sp$^2$-carbons dominating at high and low energies. The reverse is observed for the cuboidal structure, where the states with a predominant contribution of sp$^2$-carbon dominate at the Fermi region. In fact, as previously noted by Zipoli et al. \cite{Bernasconi}, $\pi$-type states are responsible for the high density of states at the Fermi level for the cuboidal structure and this is also true for the other metallic AuCu-type structure. These two 3D-cuboidal and 3D-rhombohedral polymerized structures have been proposed for the C$_{60}$ polymers synthesized at 15 GPa and 600 $^\circ$C, as referred previously \cite{yamacuboide,yamarombo}. The metallic nature of the 3D-cuboidal polymer and the semiconducting behavior of the 3D-rhombohedral polymer were experimentally observed \cite{yamacuboide,yamarombo}, although the calculated lattice parameters differ significantly from the experimental ones for both cases, as it was thoroughly discussed for the former polymer \cite{Bernasconi,tse}.

\onecolumngrid
\twocolumngrid 

The last polymerized structure being addressed in this study is the polymerized fullerite clathrate, 3D-clathrate, where most of the atoms, forty-eight out of sixty, are tetracoordinated sp$^3$-hybridized carbons. Its electronic behavior was thoroughly discussed previously \cite{LARANJEIRA2022,Laranjeira2023}, the semiconducting behavior being determined by states with a broad sp$^3$-carbon contribution that largely dominate the electronic structure, as shown in the PDOS curve given in figure \ref{pardos}. It is to be noted that hypothetical polymerized structures, based on body-centered cubic packing, with even higher sp$^3$-carbon content, having fifty-two, fifty-six and sixty sp$^3$-hybridized atoms per molecule \cite{Burgos}, are also semiconductors, thus confirming the reentrant semiconducting behavior in polymerized fullerites.

\section*{Conclusion}   

Polymerized fullerite structures present a wide variety of electronic behavior, from semiconducting to metallic. This is clearly dependent on the number of sp$^3$-hybridized carbons exhibited by each structure. The semiconducting properties are observed in structures with either high or low sp$^3$-carbon content, while the metallicity is consistently observed in structures with intermediate sp$^3$-carbon content.

The initial reduction of the bandgap with the increment in the number of sp$^3$-carbons can be interpreted as a simple consequence of the C$_{60}$ intermolecular distance reduction induced by polymeric bonding. The resulting strong interaction of the electronic densities of neighboring molecules originates this positive deformation potential. The bandgap closure is seen to be influenced by the contribution of sp$^3$-carbons, resulting from the formation of a 3D network of polymeric bonds, to the states in the Fermi region. Increasing further the number of sp$^3$-carbons leads to a bandgap reopening induced by the sp$^3$-carbon states that start to dominate the electronic structure in the Fermi region.

Finally, our study highlights this class of carbon materials as a magnificent platform for tuning electronic properties. Selecting the electronic properties could be experimentally achieved quite easily, since the polymerized structure with the sp$^3$-carbon content suitable to display a desired electronic bandgap can be selected in a simple way, just by choosing the pressure synthesis.

\begin{acknowledgments}
This work was developed within the scope of the project CICECO-Aveiro Institute of Materials, UIDB/50011/2020, UIDP/50011/2020 \& LA/P/0006/2020, financed by national funds through the FCT/MCTES (PIDDAC) and POCI-01-0145-FEDER-031326 financed through FCT and co-financed by FEDER. In addition, the 2022.07534.CEECIND researcher contract and the SFRH/BD/139327/2018 PhD grant, both funded by FCT, are gratefully acknowledged.
\end{acknowledgments}

\bibliography{referencias.bib}

\begin{thebibliography}{52}%
\makeatletter
\providecommand \@ifxundefined [1]{%
 \@ifx{#1\undefined}
}%
\providecommand \@ifnum [1]{%
 \ifnum #1\expandafter \@firstoftwo
 \else \expandafter \@secondoftwo
 \fi
}%
\providecommand \@ifx [1]{%
 \ifx #1\expandafter \@firstoftwo
 \else \expandafter \@secondoftwo
 \fi
}%
\providecommand \natexlab [1]{#1}%
\providecommand \enquote  [1]{``#1''}%
\providecommand \bibnamefont  [1]{#1}%
\providecommand \bibfnamefont [1]{#1}%
\providecommand \citenamefont [1]{#1}%
\providecommand \href@noop [0]{\@secondoftwo}%
\providecommand \href [0]{\begingroup \@sanitize@url \@href}%
\providecommand \@href[1]{\@@startlink{#1}\@@href}%
\providecommand \@@href[1]{\endgroup#1\@@endlink}%
\providecommand \@sanitize@url [0]{\catcode `\\12\catcode `\$12\catcode `\&12\catcode `\#12\catcode `\^12\catcode `\_12\catcode `\%12\relax}%
\providecommand \@@startlink[1]{}%
\providecommand \@@endlink[0]{}%
\providecommand \url  [0]{\begingroup\@sanitize@url \@url }%
\providecommand \@url [1]{\endgroup\@href {#1}{\urlprefix }}%
\providecommand \urlprefix  [0]{URL }%
\providecommand \Eprint [0]{\href }%
\providecommand \doibase [0]{https://doi.org/}%
\providecommand \selectlanguage [0]{\@gobble}%
\providecommand \bibinfo  [0]{\@secondoftwo}%
\providecommand \bibfield  [0]{\@secondoftwo}%
\providecommand \translation [1]{[#1]}%
\providecommand \BibitemOpen [0]{}%
\providecommand \bibitemStop [0]{}%
\providecommand \bibitemNoStop [0]{.\EOS\space}%
\providecommand \EOS [0]{\spacefactor3000\relax}%
\providecommand \BibitemShut  [1]{\csname bibitem#1\endcsname}%
\let\auto@bib@innerbib\@empty
\bibitem [{\citenamefont {Peplow}(2013)}]{peplow2013graphene_cit1}%
  \BibitemOpen
  \bibfield  {author} {\bibinfo {author} {\bibfnamefont {M.}~\bibnamefont {Peplow}},\ }\href@noop {} {\bibfield  {journal} {\bibinfo  {journal} {Nature}\ }\textbf {\bibinfo {volume} {503}},\ \bibinfo {pages} {327} (\bibinfo {year} {2013})}\BibitemShut {NoStop}%
\bibitem [{\citenamefont {Kang}\ \emph {et~al.}(2017)\citenamefont {Kang}, \citenamefont {Kim}, \citenamefont {Ryu}, \citenamefont {Jung}, \citenamefont {Kim}, \citenamefont {Moreschini}, \citenamefont {Jozwiak}, \citenamefont {Rotenberg}, \citenamefont {Bostwick},\ and\ \citenamefont {Kim}}]{bandgap1}%
  \BibitemOpen
  \bibfield  {author} {\bibinfo {author} {\bibfnamefont {M.}~\bibnamefont {Kang}}, \bibinfo {author} {\bibfnamefont {B.}~\bibnamefont {Kim}}, \bibinfo {author} {\bibfnamefont {S.}~\bibnamefont {Ryu}}, \bibinfo {author} {\bibfnamefont {S.}~\bibnamefont {Jung}}, \bibinfo {author} {\bibfnamefont {J.}~\bibnamefont {Kim}}, \bibinfo {author} {\bibfnamefont {L.}~\bibnamefont {Moreschini}}, \bibinfo {author} {\bibfnamefont {C.}~\bibnamefont {Jozwiak}}, \bibinfo {author} {\bibfnamefont {E.}~\bibnamefont {Rotenberg}}, \bibinfo {author} {\bibfnamefont {A.}~\bibnamefont {Bostwick}},\ and\ \bibinfo {author} {\bibfnamefont {K.}~\bibnamefont {Kim}},\ }\href {https://doi.org/10.1021/acs.nanolett.6b04775} {\bibfield  {journal} {\bibinfo  {journal} {Nano Lett.}\ }\textbf {\bibinfo {volume} {17}},\ \bibinfo {pages} {1610} (\bibinfo {year} {2017})},\ \bibinfo {note} {pMID: 28118710}\BibitemShut {NoStop}%
\bibitem [{\citenamefont {Scuseria}(2021)}]{bandgap_scuseria}%
  \BibitemOpen
  \bibfield  {author} {\bibinfo {author} {\bibfnamefont {G.~E.}\ \bibnamefont {Scuseria}},\ }\href {https://doi.org/10.1073/pnas.2113648118} {\bibfield  {journal} {\bibinfo  {journal} {Proceedings of the National Academy of Sciences}\ }\textbf {\bibinfo {volume} {118}},\ \bibinfo {pages} {e2113648118} (\bibinfo {year} {2021})},\ \Eprint {https://arxiv.org/abs/https://www.pnas.org/doi/pdf/10.1073/pnas.2113648118} {https://www.pnas.org/doi/pdf/10.1073/pnas.2113648118} \BibitemShut {NoStop}%
\bibitem [{\citenamefont {Capasso}(1987)}]{capasso1987band_cit2}%
  \BibitemOpen
  \bibfield  {author} {\bibinfo {author} {\bibfnamefont {F.}~\bibnamefont {Capasso}},\ }\href@noop {} {\bibfield  {journal} {\bibinfo  {journal} {Science}\ }\textbf {\bibinfo {volume} {235}},\ \bibinfo {pages} {172} (\bibinfo {year} {1987})}\BibitemShut {NoStop}%
\bibitem [{\citenamefont {Zeghbroeck}(2007)}]{zeghbroeck2007principles_cit3}%
  \BibitemOpen
  \bibfield  {author} {\bibinfo {author} {\bibfnamefont {V.}~\bibnamefont {Zeghbroeck}},\ }\href {https://books.google.pt/books?id=8NkXkgEACAAJ} {\emph {\bibinfo {title} {Principles of Semiconductor Devices and Heterojunctions}}}\ (\bibinfo  {publisher} {Prentice Hall PTR},\ \bibinfo {year} {2007})\BibitemShut {NoStop}%
\bibitem [{\citenamefont {Trickey}\ \emph {et~al.}(1992)\citenamefont {Trickey}, \citenamefont {M\"uller-Plathe}, \citenamefont {Diercksen},\ and\ \citenamefont {Boettger}}]{graphene_old}%
  \BibitemOpen
  \bibfield  {author} {\bibinfo {author} {\bibfnamefont {S.}~\bibnamefont {Trickey}}, \bibinfo {author} {\bibfnamefont {F.}~\bibnamefont {M\"uller-Plathe}}, \bibinfo {author} {\bibfnamefont {G.}~\bibnamefont {Diercksen}},\ and\ \bibinfo {author} {\bibfnamefont {J.}~\bibnamefont {Boettger}},\ }\href {https://doi.org/10.1103/PhysRevB.45.4460} {\bibfield  {journal} {\bibinfo  {journal} {Phys. Rev. B}\ }\textbf {\bibinfo {volume} {45}},\ \bibinfo {pages} {4460} (\bibinfo {year} {1992})}\BibitemShut {NoStop}%
\bibitem [{\citenamefont {Partoens}\ and\ \citenamefont {Peeters}(2006)}]{graphite_tb}%
  \BibitemOpen
  \bibfield  {author} {\bibinfo {author} {\bibfnamefont {B.}~\bibnamefont {Partoens}}\ and\ \bibinfo {author} {\bibfnamefont {F.}~\bibnamefont {Peeters}},\ }\href {https://doi.org/10.1103/PhysRevB.74.075404} {\bibfield  {journal} {\bibinfo  {journal} {Phys. Rev. B}\ }\textbf {\bibinfo {volume} {74}},\ \bibinfo {pages} {075404} (\bibinfo {year} {2006})}\BibitemShut {NoStop}%
\bibitem [{\citenamefont {Strong}\ \emph {et~al.}(2004)\citenamefont {Strong}, \citenamefont {Pickard}, \citenamefont {Milman}, \citenamefont {Thimm},\ and\ \citenamefont {Winkler}}]{dgap}%
  \BibitemOpen
  \bibfield  {author} {\bibinfo {author} {\bibfnamefont {R.}~\bibnamefont {Strong}}, \bibinfo {author} {\bibfnamefont {C.}~\bibnamefont {Pickard}}, \bibinfo {author} {\bibfnamefont {V.}~\bibnamefont {Milman}}, \bibinfo {author} {\bibfnamefont {G.}~\bibnamefont {Thimm}},\ and\ \bibinfo {author} {\bibfnamefont {B.}~\bibnamefont {Winkler}},\ }\href {https://doi.org/10.1103/PhysRevB.70.045101} {\bibfield  {journal} {\bibinfo  {journal} {Phys. Rev. B}\ }\textbf {\bibinfo {volume} {70}},\ \bibinfo {pages} {045101} (\bibinfo {year} {2004})}\BibitemShut {NoStop}%
\bibitem [{\citenamefont {Otani}\ \emph {et~al.}(2003)\citenamefont {Otani}, \citenamefont {Okada},\ and\ \citenamefont {Oshiyama}}]{encap_enano}%
  \BibitemOpen
  \bibfield  {author} {\bibinfo {author} {\bibfnamefont {M.}~\bibnamefont {Otani}}, \bibinfo {author} {\bibfnamefont {S.}~\bibnamefont {Okada}},\ and\ \bibinfo {author} {\bibfnamefont {A.}~\bibnamefont {Oshiyama}},\ }\href {https://doi.org/10.1103/PhysRevB.68.125424} {\bibfield  {journal} {\bibinfo  {journal} {Phys. Rev. B}\ }\textbf {\bibinfo {volume} {68}},\ \bibinfo {pages} {125424} (\bibinfo {year} {2003})}\BibitemShut {NoStop}%
\bibitem [{\citenamefont {An}\ and\ \citenamefont {Lee}(2006)}]{notubeselectronic}%
  \BibitemOpen
  \bibfield  {author} {\bibinfo {author} {\bibfnamefont {K.}~\bibnamefont {An}}\ and\ \bibinfo {author} {\bibfnamefont {Y.}~\bibnamefont {Lee}},\ }\href@noop {} {\bibfield  {journal} {\bibinfo  {journal} {Nano}\ }\textbf {\bibinfo {volume} {1}},\ \bibinfo {pages} {115} (\bibinfo {year} {2006})}\BibitemShut {NoStop}%
\bibitem [{\citenamefont {Barone}\ \emph {et~al.}(2006)\citenamefont {Barone}, \citenamefont {Hod},\ and\ \citenamefont {Scuseria}}]{nanoribons_scuseria}%
  \BibitemOpen
  \bibfield  {author} {\bibinfo {author} {\bibfnamefont {V.}~\bibnamefont {Barone}}, \bibinfo {author} {\bibfnamefont {O.}~\bibnamefont {Hod}},\ and\ \bibinfo {author} {\bibfnamefont {G.}~\bibnamefont {Scuseria}},\ }\href {https://doi.org/10.1021/nl0617033} {\bibfield  {journal} {\bibinfo  {journal} {Nano Lett.}\ }\textbf {\bibinfo {volume} {6}},\ \bibinfo {pages} {2748} (\bibinfo {year} {2006})}\BibitemShut {NoStop}%
\bibitem [{\citenamefont {Cao}\ \emph {et~al.}(2018)\citenamefont {Cao}, \citenamefont {Fatemi}, \citenamefont {Fang}, \citenamefont {Watanabe}, \citenamefont {Taniguchi}, \citenamefont {Kaxiras},\ and\ \citenamefont {Jarillo-Herrero}}]{cao2018unconventional}%
  \BibitemOpen
  \bibfield  {author} {\bibinfo {author} {\bibfnamefont {Y.}~\bibnamefont {Cao}}, \bibinfo {author} {\bibfnamefont {V.}~\bibnamefont {Fatemi}}, \bibinfo {author} {\bibfnamefont {S.}~\bibnamefont {Fang}}, \bibinfo {author} {\bibfnamefont {K.}~\bibnamefont {Watanabe}}, \bibinfo {author} {\bibfnamefont {T.}~\bibnamefont {Taniguchi}}, \bibinfo {author} {\bibfnamefont {E.}~\bibnamefont {Kaxiras}},\ and\ \bibinfo {author} {\bibfnamefont {P.}~\bibnamefont {Jarillo-Herrero}},\ }\href@noop {} {\bibfield  {journal} {\bibinfo  {journal} {Nature}\ }\textbf {\bibinfo {volume} {556}},\ \bibinfo {pages} {43} (\bibinfo {year} {2018})}\BibitemShut {NoStop}%
\bibitem [{\citenamefont {Oh}\ \emph {et~al.}(2021)\citenamefont {Oh}, \citenamefont {Nuckolls}, \citenamefont {Wong}, \citenamefont {Lee}, \citenamefont {Liu}, \citenamefont {Watanabe}, \citenamefont {Taniguchi},\ and\ \citenamefont {Yazdani}}]{oh2021evidence}%
  \BibitemOpen
  \bibfield  {author} {\bibinfo {author} {\bibfnamefont {M.}~\bibnamefont {Oh}}, \bibinfo {author} {\bibfnamefont {K.~P.}\ \bibnamefont {Nuckolls}}, \bibinfo {author} {\bibfnamefont {D.}~\bibnamefont {Wong}}, \bibinfo {author} {\bibfnamefont {R.~L.}\ \bibnamefont {Lee}}, \bibinfo {author} {\bibfnamefont {X.}~\bibnamefont {Liu}}, \bibinfo {author} {\bibfnamefont {K.}~\bibnamefont {Watanabe}}, \bibinfo {author} {\bibfnamefont {T.}~\bibnamefont {Taniguchi}},\ and\ \bibinfo {author} {\bibfnamefont {A.}~\bibnamefont {Yazdani}},\ }\href@noop {} {\bibfield  {journal} {\bibinfo  {journal} {Nature}\ }\textbf {\bibinfo {volume} {600}},\ \bibinfo {pages} {240} (\bibinfo {year} {2021})}\BibitemShut {NoStop}%
\bibitem [{\citenamefont {Kim}\ \emph {et~al.}(2022)\citenamefont {Kim}, \citenamefont {Choi}, \citenamefont {Lewandowski}, \citenamefont {Thomson}, \citenamefont {Zhang}, \citenamefont {Polski}, \citenamefont {Watanabe}, \citenamefont {Taniguchi}, \citenamefont {Alicea},\ and\ \citenamefont {Nadj-Perge}}]{kim2022evidence}%
  \BibitemOpen
  \bibfield  {author} {\bibinfo {author} {\bibfnamefont {H.}~\bibnamefont {Kim}}, \bibinfo {author} {\bibfnamefont {Y.}~\bibnamefont {Choi}}, \bibinfo {author} {\bibfnamefont {C.}~\bibnamefont {Lewandowski}}, \bibinfo {author} {\bibfnamefont {A.}~\bibnamefont {Thomson}}, \bibinfo {author} {\bibfnamefont {Y.}~\bibnamefont {Zhang}}, \bibinfo {author} {\bibfnamefont {R.}~\bibnamefont {Polski}}, \bibinfo {author} {\bibfnamefont {K.}~\bibnamefont {Watanabe}}, \bibinfo {author} {\bibfnamefont {T.}~\bibnamefont {Taniguchi}}, \bibinfo {author} {\bibfnamefont {J.}~\bibnamefont {Alicea}},\ and\ \bibinfo {author} {\bibfnamefont {S.}~\bibnamefont {Nadj-Perge}},\ }\href@noop {} {\bibfield  {journal} {\bibinfo  {journal} {Nature}\ }\textbf {\bibinfo {volume} {606}},\ \bibinfo {pages} {494} (\bibinfo {year} {2022})}\BibitemShut {NoStop}%
\bibitem [{\citenamefont {Saito}\ \emph {et~al.}(1992)\citenamefont {Saito}, \citenamefont {Shinohara}, \citenamefont {Kato}, \citenamefont {Nagashima}, \citenamefont {Ohkohchi},\ and\ \citenamefont {Ando}}]{saito1992electric}%
  \BibitemOpen
  \bibfield  {author} {\bibinfo {author} {\bibfnamefont {Y.}~\bibnamefont {Saito}}, \bibinfo {author} {\bibfnamefont {H.}~\bibnamefont {Shinohara}}, \bibinfo {author} {\bibfnamefont {M.}~\bibnamefont {Kato}}, \bibinfo {author} {\bibfnamefont {H.}~\bibnamefont {Nagashima}}, \bibinfo {author} {\bibfnamefont {M.}~\bibnamefont {Ohkohchi}},\ and\ \bibinfo {author} {\bibfnamefont {Y.}~\bibnamefont {Ando}},\ }\href@noop {} {\bibfield  {journal} {\bibinfo  {journal} {Chem. Phys. Lett.}\ }\textbf {\bibinfo {volume} {189}},\ \bibinfo {pages} {236} (\bibinfo {year} {1992})}\BibitemShut {NoStop}%
\bibitem [{\citenamefont {Qiu}\ \emph {et~al.}(2006)\citenamefont {Qiu}, \citenamefont {Chowdhury}, \citenamefont {Hammer}, \citenamefont {Velisavljevic}, \citenamefont {Baker},\ and\ \citenamefont {Vohra}}]{Vohra_gap}%
  \BibitemOpen
  \bibfield  {author} {\bibinfo {author} {\bibfnamefont {W.}~\bibnamefont {Qiu}}, \bibinfo {author} {\bibfnamefont {S.}~\bibnamefont {Chowdhury}}, \bibinfo {author} {\bibfnamefont {R.}~\bibnamefont {Hammer}}, \bibinfo {author} {\bibfnamefont {N.}~\bibnamefont {Velisavljevic}}, \bibinfo {author} {\bibfnamefont {P.}~\bibnamefont {Baker}},\ and\ \bibinfo {author} {\bibfnamefont {Y.}~\bibnamefont {Vohra}},\ }\href@noop {} {\bibfield  {journal} {\bibinfo  {journal} {High Press. Res.}\ }\textbf {\bibinfo {volume} {26}},\ \bibinfo {pages} {175} (\bibinfo {year} {2006})}\BibitemShut {NoStop}%
\bibitem [{\citenamefont {Moshary}\ \emph {et~al.}(1992)\citenamefont {Moshary}, \citenamefont {Chen}, \citenamefont {Silvera}, \citenamefont {Brown}, \citenamefont {Dorn}, \citenamefont {de~Vries},\ and\ \citenamefont {Bethune}}]{moshary1992gap}%
  \BibitemOpen
  \bibfield  {author} {\bibinfo {author} {\bibfnamefont {F.}~\bibnamefont {Moshary}}, \bibinfo {author} {\bibfnamefont {N.}~\bibnamefont {Chen}}, \bibinfo {author} {\bibfnamefont {I.}~\bibnamefont {Silvera}}, \bibinfo {author} {\bibfnamefont {C.}~\bibnamefont {Brown}}, \bibinfo {author} {\bibfnamefont {H.}~\bibnamefont {Dorn}}, \bibinfo {author} {\bibfnamefont {M.}~\bibnamefont {de~Vries}},\ and\ \bibinfo {author} {\bibfnamefont {D.}~\bibnamefont {Bethune}},\ }\href@noop {} {\bibfield  {journal} {\bibinfo  {journal} {Phys. Rev. Lett.}\ }\textbf {\bibinfo {volume} {69}},\ \bibinfo {pages} {466} (\bibinfo {year} {1992})}\BibitemShut {NoStop}%
\bibitem [{\citenamefont {N\'u\~nez$\mathrm{-}$Regueiro}\ \emph {et~al.}(1991)\citenamefont {N\'u\~nez$\mathrm{-}$Regueiro}, \citenamefont {Monceau}, \citenamefont {Rassat}, \citenamefont {Bernier},\ and\ \citenamefont {Zahab}}]{regueiro1991gap}%
  \BibitemOpen
  \bibfield  {author} {\bibinfo {author} {\bibfnamefont {M.}~\bibnamefont {N\'u\~nez$\mathrm{-}$Regueiro}}, \bibinfo {author} {\bibfnamefont {P.}~\bibnamefont {Monceau}}, \bibinfo {author} {\bibfnamefont {A.}~\bibnamefont {Rassat}}, \bibinfo {author} {\bibfnamefont {P.}~\bibnamefont {Bernier}},\ and\ \bibinfo {author} {\bibfnamefont {A.}~\bibnamefont {Zahab}},\ }\href@noop {} {\bibfield  {journal} {\bibinfo  {journal} {Nature}\ }\textbf {\bibinfo {volume} {354}},\ \bibinfo {pages} {289} (\bibinfo {year} {1991})}\BibitemShut {NoStop}%
\bibitem [{\citenamefont {Wu}\ \emph {et~al.}(2022)\citenamefont {Wu}, \citenamefont {Gao}, \citenamefont {Zhang}, \citenamefont {Soldatov}, \citenamefont {Kim}, \citenamefont {Wang},\ and\ \citenamefont {Tian}}]{c60_xlynes}%
  \BibitemOpen
  \bibfield  {author} {\bibinfo {author} {\bibfnamefont {Z.}~\bibnamefont {Wu}}, \bibinfo {author} {\bibfnamefont {G.}~\bibnamefont {Gao}}, \bibinfo {author} {\bibfnamefont {J.}~\bibnamefont {Zhang}}, \bibinfo {author} {\bibfnamefont {A.}~\bibnamefont {Soldatov}}, \bibinfo {author} {\bibfnamefont {J.}~\bibnamefont {Kim}}, \bibinfo {author} {\bibfnamefont {L.}~\bibnamefont {Wang}},\ and\ \bibinfo {author} {\bibfnamefont {Y.}~\bibnamefont {Tian}},\ }\href@noop {} {\bibfield  {journal} {\bibinfo  {journal} {Nano Res.}\ }\textbf {\bibinfo {volume} {15}},\ \bibinfo {pages} {3788} (\bibinfo {year} {2022})}\BibitemShut {NoStop}%
\bibitem [{\citenamefont {Meirzadeh}\ \emph {et~al.}(2023)\citenamefont {Meirzadeh}, \citenamefont {Evans}, \citenamefont {Rezaee}, \citenamefont {Milich}, \citenamefont {Dionne}, \citenamefont {Darlington}, \citenamefont {Bao}, \citenamefont {Bartholomew}, \citenamefont {Handa}, \citenamefont {Rizzo}, \citenamefont {Wiscons}, \citenamefont {Reza}, \citenamefont {Zangiabadi}, \citenamefont {Fardian-Melamed}, \citenamefont {Crowther}, \citenamefont {Schuck}, \citenamefont {Basov}, \citenamefont {Zhu}, \citenamefont {Giri}, \citenamefont {Hopkins}, \citenamefont {Kim}, \citenamefont {Steigerwald}, \citenamefont {Yang}, \citenamefont {Nuckolls},\ and\ \citenamefont {Roy}}]{naturemgc60_23}%
  \BibitemOpen
  \bibfield  {author} {\bibinfo {author} {\bibfnamefont {E.}~\bibnamefont {Meirzadeh}}, \bibinfo {author} {\bibfnamefont {A.}~\bibnamefont {Evans}}, \bibinfo {author} {\bibfnamefont {M.}~\bibnamefont {Rezaee}}, \bibinfo {author} {\bibfnamefont {M.}~\bibnamefont {Milich}}, \bibinfo {author} {\bibfnamefont {C.}~\bibnamefont {Dionne}}, \bibinfo {author} {\bibfnamefont {T.}~\bibnamefont {Darlington}}, \bibinfo {author} {\bibfnamefont {S.~T.}\ \bibnamefont {Bao}}, \bibinfo {author} {\bibfnamefont {A.}~\bibnamefont {Bartholomew}}, \bibinfo {author} {\bibfnamefont {T.}~\bibnamefont {Handa}}, \bibinfo {author} {\bibfnamefont {D.}~\bibnamefont {Rizzo}}, \bibinfo {author} {\bibfnamefont {R.}~\bibnamefont {Wiscons}}, \bibinfo {author} {\bibfnamefont {M.}~\bibnamefont {Reza}}, \bibinfo {author} {\bibfnamefont {A.}~\bibnamefont {Zangiabadi}}, \bibinfo {author} {\bibfnamefont {N.}~\bibnamefont {Fardian-Melamed}}, \bibinfo {author} {\bibfnamefont {A.}~\bibnamefont {Crowther}}, \bibinfo {author} {\bibfnamefont
  {J.}~\bibnamefont {Schuck}}, \bibinfo {author} {\bibfnamefont {D.}~\bibnamefont {Basov}}, \bibinfo {author} {\bibfnamefont {X.}~\bibnamefont {Zhu}}, \bibinfo {author} {\bibfnamefont {A.}~\bibnamefont {Giri}}, \bibinfo {author} {\bibfnamefont {P.}~\bibnamefont {Hopkins}}, \bibinfo {author} {\bibfnamefont {P.}~\bibnamefont {Kim}}, \bibinfo {author} {\bibfnamefont {M.}~\bibnamefont {Steigerwald}}, \bibinfo {author} {\bibfnamefont {J.}~\bibnamefont {Yang}}, \bibinfo {author} {\bibfnamefont {C.}~\bibnamefont {Nuckolls}},\ and\ \bibinfo {author} {\bibfnamefont {X.}~\bibnamefont {Roy}},\ }\href@noop {} {\bibfield  {journal} {\bibinfo  {journal} {Nature}\ }\textbf {\bibinfo {volume} {613}},\ \bibinfo {pages} {71} (\bibinfo {year} {2023})}\BibitemShut {NoStop}%
\bibitem [{\citenamefont {Hou}\ \emph {et~al.}(2022)\citenamefont {Hou}, \citenamefont {Cui}, \citenamefont {Guan}, \citenamefont {Wang}, \citenamefont {Li}, \citenamefont {Liu}, \citenamefont {Zhu},\ and\ \citenamefont {Zheng}}]{naturemgc60_22}%
  \BibitemOpen
  \bibfield  {author} {\bibinfo {author} {\bibfnamefont {L.}~\bibnamefont {Hou}}, \bibinfo {author} {\bibfnamefont {X.}~\bibnamefont {Cui}}, \bibinfo {author} {\bibfnamefont {B.}~\bibnamefont {Guan}}, \bibinfo {author} {\bibfnamefont {S.}~\bibnamefont {Wang}}, \bibinfo {author} {\bibfnamefont {R.}~\bibnamefont {Li}}, \bibinfo {author} {\bibfnamefont {Y.}~\bibnamefont {Liu}}, \bibinfo {author} {\bibfnamefont {D.}~\bibnamefont {Zhu}},\ and\ \bibinfo {author} {\bibfnamefont {J.}~\bibnamefont {Zheng}},\ }\href@noop {} {\bibfield  {journal} {\bibinfo  {journal} {Nature}\ }\textbf {\bibinfo {volume} {606}},\ \bibinfo {pages} {507} (\bibinfo {year} {2022})}\BibitemShut {NoStop}%
\bibitem [{\citenamefont {Pan}\ \emph {et~al.}(2023)\citenamefont {Pan}, \citenamefont {Ni}, \citenamefont {Xu}, \citenamefont {Chen}, \citenamefont {Wang}, \citenamefont {Gong}, \citenamefont {Liu}, \citenamefont {Li}, \citenamefont {Lin}, \citenamefont {Li}, \citenamefont {Wang}, \citenamefont {Yan}, \citenamefont {Yin}, \citenamefont {Tan}, \citenamefont {Sun}, \citenamefont {Yu}, \citenamefont {Ruoff},\ and\ \citenamefont {Zhu}}]{pan2023long}%
  \BibitemOpen
  \bibfield  {author} {\bibinfo {author} {\bibfnamefont {F.}~\bibnamefont {Pan}}, \bibinfo {author} {\bibfnamefont {K.}~\bibnamefont {Ni}}, \bibinfo {author} {\bibfnamefont {T.}~\bibnamefont {Xu}}, \bibinfo {author} {\bibfnamefont {H.}~\bibnamefont {Chen}}, \bibinfo {author} {\bibfnamefont {Y.}~\bibnamefont {Wang}}, \bibinfo {author} {\bibfnamefont {K.}~\bibnamefont {Gong}}, \bibinfo {author} {\bibfnamefont {C.}~\bibnamefont {Liu}}, \bibinfo {author} {\bibfnamefont {X.}~\bibnamefont {Li}}, \bibinfo {author} {\bibfnamefont {M.-L.}\ \bibnamefont {Lin}}, \bibinfo {author} {\bibfnamefont {S.}~\bibnamefont {Li}}, \bibinfo {author} {\bibfnamefont {X.}~\bibnamefont {Wang}}, \bibinfo {author} {\bibfnamefont {W.}~\bibnamefont {Yan}}, \bibinfo {author} {\bibfnamefont {W.}~\bibnamefont {Yin}}, \bibinfo {author} {\bibfnamefont {P.-H.}\ \bibnamefont {Tan}}, \bibinfo {author} {\bibfnamefont {L.}~\bibnamefont {Sun}}, \bibinfo {author} {\bibfnamefont {D.}~\bibnamefont {Yu}}, \bibinfo {author} {\bibfnamefont {R.}~\bibnamefont
  {Ruoff}},\ and\ \bibinfo {author} {\bibfnamefont {Y.}~\bibnamefont {Zhu}},\ }\href@noop {} {\bibfield  {journal} {\bibinfo  {journal} {Nature}\ }\textbf {\bibinfo {volume} {614}},\ \bibinfo {pages} {95} (\bibinfo {year} {2023})}\BibitemShut {NoStop}%
\bibitem [{\citenamefont {Wang}\ \emph {et~al.}(2023)\citenamefont {Wang}, \citenamefont {Zhang}, \citenamefont {Wu}, \citenamefont {Chen}, \citenamefont {Yang}, \citenamefont {Lu},\ and\ \citenamefont {Du}}]{wang2023}%
  \BibitemOpen
  \bibfield  {author} {\bibinfo {author} {\bibfnamefont {T.}~\bibnamefont {Wang}}, \bibinfo {author} {\bibfnamefont {L.}~\bibnamefont {Zhang}}, \bibinfo {author} {\bibfnamefont {J.}~\bibnamefont {Wu}}, \bibinfo {author} {\bibfnamefont {M.}~\bibnamefont {Chen}}, \bibinfo {author} {\bibfnamefont {S.}~\bibnamefont {Yang}}, \bibinfo {author} {\bibfnamefont {Y.}~\bibnamefont {Lu}},\ and\ \bibinfo {author} {\bibfnamefont {P.}~\bibnamefont {Du}},\ }\href {https://doi.org/https://doi.org/10.1002/anie.202311352} {\bibfield  {journal} {\bibinfo  {journal} {Angewandte Chemie International Edition}\ }\textbf {\bibinfo {volume} {62}},\ \bibinfo {pages} {e202311352} (\bibinfo {year} {2023})}\BibitemShut {NoStop}%
\bibitem [{\citenamefont {N\'u\~nez$\mathrm{-}$Regueiro}\ \emph {et~al.}(1995)\citenamefont {N\'u\~nez$\mathrm{-}$Regueiro}, \citenamefont {Marques}, \citenamefont {Hodeau}, \citenamefont {B\'ethoux},\ and\ \citenamefont {Perroux}}]{marques_prl}%
  \BibitemOpen
  \bibfield  {author} {\bibinfo {author} {\bibfnamefont {M.}~\bibnamefont {N\'u\~nez$\mathrm{-}$Regueiro}}, \bibinfo {author} {\bibfnamefont {L.}~\bibnamefont {Marques}}, \bibinfo {author} {\bibfnamefont {J.-L.}\ \bibnamefont {Hodeau}}, \bibinfo {author} {\bibfnamefont {O.}~\bibnamefont {B\'ethoux}},\ and\ \bibinfo {author} {\bibfnamefont {M.}~\bibnamefont {Perroux}},\ }\href {https://doi.org/10.1103/PhysRevLett.74.278} {\bibfield  {journal} {\bibinfo  {journal} {Phys. Rev. Lett.}\ }\textbf {\bibinfo {volume} {74}},\ \bibinfo {pages} {278} (\bibinfo {year} {1995})}\BibitemShut {NoStop}%
\bibitem [{\citenamefont {Álvarez$\mathrm{-}$Murga}\ and\ \citenamefont {Hodeau}(2015)}]{alvarez}%
  \BibitemOpen
  \bibfield  {author} {\bibinfo {author} {\bibfnamefont {M.}~\bibnamefont {Álvarez$\mathrm{-}$Murga}}\ and\ \bibinfo {author} {\bibfnamefont {J.-L.}\ \bibnamefont {Hodeau}},\ }\href {https://doi.org/http://dx.doi.org/10.1016/j.carbon.2014.10.083} {\bibfield  {journal} {\bibinfo  {journal} {Carbon}\ }\textbf {\bibinfo {volume} {82}},\ \bibinfo {pages} {381 } (\bibinfo {year} {2015})}\BibitemShut {NoStop}%
\bibitem [{\citenamefont {Laranjeira}\ \emph {et~al.}(2017)\citenamefont {Laranjeira}, \citenamefont {Marques}, \citenamefont {Mezouar}, \citenamefont {Melle-Franco},\ and\ \citenamefont {Strutyński}}]{Laranjeira2017}%
  \BibitemOpen
  \bibfield  {author} {\bibinfo {author} {\bibfnamefont {J.}~\bibnamefont {Laranjeira}}, \bibinfo {author} {\bibfnamefont {L.}~\bibnamefont {Marques}}, \bibinfo {author} {\bibfnamefont {M.}~\bibnamefont {Mezouar}}, \bibinfo {author} {\bibfnamefont {M.}~\bibnamefont {Melle-Franco}},\ and\ \bibinfo {author} {\bibfnamefont {K.}~\bibnamefont {Strutyński}},\ }\href {https://doi.org/10.1002/pssr.201700343} {\bibfield  {journal} {\bibinfo  {journal} {Phys. status solidi -RRL}\ }\textbf {\bibinfo {volume} {11}},\ \bibinfo {pages} {1700343} (\bibinfo {year} {2017})}\BibitemShut {NoStop}%
\bibitem [{\citenamefont {Laranjeira}\ \emph {et~al.}(2019)\citenamefont {Laranjeira}, \citenamefont {Marques}, \citenamefont {Mezouar}, \citenamefont {Melle-Franco},\ and\ \citenamefont {Strutyński}}]{Laranjeira2019}%
  \BibitemOpen
  \bibfield  {author} {\bibinfo {author} {\bibfnamefont {J.}~\bibnamefont {Laranjeira}}, \bibinfo {author} {\bibfnamefont {L.}~\bibnamefont {Marques}}, \bibinfo {author} {\bibfnamefont {M.}~\bibnamefont {Mezouar}}, \bibinfo {author} {\bibfnamefont {M.}~\bibnamefont {Melle-Franco}},\ and\ \bibinfo {author} {\bibfnamefont {K.}~\bibnamefont {Strutyński}},\ }\href {https://doi.org/https://doi.org/10.1016/j.mlblux.2019.100026} {\bibfield  {journal} {\bibinfo  {journal} {Mater. Lett.: X}\ }\textbf {\bibinfo {volume} {4}},\ \bibinfo {pages} {100026} (\bibinfo {year} {2019})}\BibitemShut {NoStop}%
\bibitem [{\citenamefont {Marques}\ \emph {et~al.}(1999)\citenamefont {Marques}, \citenamefont {Mezouar}, \citenamefont {Hodeau}, \citenamefont {N{\'u}{\~n}ez-Regueiro}, \citenamefont {Serebryanaya}, \citenamefont {Ivdenko}, \citenamefont {Blank},\ and\ \citenamefont {Dubitsky}}]{marquesscience}%
  \BibitemOpen
  \bibfield  {author} {\bibinfo {author} {\bibfnamefont {L.}~\bibnamefont {Marques}}, \bibinfo {author} {\bibfnamefont {M.}~\bibnamefont {Mezouar}}, \bibinfo {author} {\bibfnamefont {J.-L.}\ \bibnamefont {Hodeau}}, \bibinfo {author} {\bibfnamefont {M.}~\bibnamefont {N{\'u}{\~n}ez-Regueiro}}, \bibinfo {author} {\bibfnamefont {N.}~\bibnamefont {Serebryanaya}}, \bibinfo {author} {\bibfnamefont {V.}~\bibnamefont {Ivdenko}}, \bibinfo {author} {\bibfnamefont {V.}~\bibnamefont {Blank}},\ and\ \bibinfo {author} {\bibfnamefont {G.}~\bibnamefont {Dubitsky}},\ }\href {https://doi.org/10.1126/science.283.5408.1720} {\bibfield  {journal} {\bibinfo  {journal} {Science}\ }\textbf {\bibinfo {volume} {283}},\ \bibinfo {pages} {1720} (\bibinfo {year} {1999})}\BibitemShut {NoStop}%
\bibitem [{\citenamefont {Yamanaka}\ \emph {et~al.}(2006)\citenamefont {Yamanaka}, \citenamefont {Kubo}, \citenamefont {Inumaru}, \citenamefont {Komaguchi}, \citenamefont {Kini}, \citenamefont {Inoue},\ and\ \citenamefont {Irifune}}]{yamacuboide}%
  \BibitemOpen
  \bibfield  {author} {\bibinfo {author} {\bibfnamefont {S.}~\bibnamefont {Yamanaka}}, \bibinfo {author} {\bibfnamefont {A.}~\bibnamefont {Kubo}}, \bibinfo {author} {\bibfnamefont {K.}~\bibnamefont {Inumaru}}, \bibinfo {author} {\bibfnamefont {K.}~\bibnamefont {Komaguchi}}, \bibinfo {author} {\bibfnamefont {N.}~\bibnamefont {Kini}}, \bibinfo {author} {\bibfnamefont {T.}~\bibnamefont {Inoue}},\ and\ \bibinfo {author} {\bibfnamefont {T.}~\bibnamefont {Irifune}},\ }\href {https://doi.org/10.1103/PhysRevLett.96.076602} {\bibfield  {journal} {\bibinfo  {journal} {Phys. Rev. Lett.}\ }\textbf {\bibinfo {volume} {96}},\ \bibinfo {pages} {076602} (\bibinfo {year} {2006})}\BibitemShut {NoStop}%
\bibitem [{\citenamefont {Yamanaka}\ \emph {et~al.}(2008)\citenamefont {Yamanaka}, \citenamefont {Kini}, \citenamefont {Kubo}, \citenamefont {Jida},\ and\ \citenamefont {Kuramoto}}]{yamarombo}%
  \BibitemOpen
  \bibfield  {author} {\bibinfo {author} {\bibfnamefont {S.}~\bibnamefont {Yamanaka}}, \bibinfo {author} {\bibfnamefont {N.}~\bibnamefont {Kini}}, \bibinfo {author} {\bibfnamefont {A.}~\bibnamefont {Kubo}}, \bibinfo {author} {\bibfnamefont {S.}~\bibnamefont {Jida}},\ and\ \bibinfo {author} {\bibfnamefont {H.}~\bibnamefont {Kuramoto}},\ }\href {https://doi.org/10.1021/ja076761k} {\bibfield  {journal} {\bibinfo  {journal} {J. Am. Chem. Soc.}\ }\textbf {\bibinfo {volume} {130}},\ \bibinfo {pages} {4303} (\bibinfo {year} {2008})}\BibitemShut {NoStop}%
\bibitem [{\citenamefont {Blank}\ \emph {et~al.}(1998)\citenamefont {Blank}, \citenamefont {Buga}, \citenamefont {Dubitsky}, \citenamefont {{Serebryanaya}}, \citenamefont {Popov},\ and\ \citenamefont {Sundqvist}}]{BLANK1998}%
  \BibitemOpen
  \bibfield  {author} {\bibinfo {author} {\bibfnamefont {V.}~\bibnamefont {Blank}}, \bibinfo {author} {\bibfnamefont {S.}~\bibnamefont {Buga}}, \bibinfo {author} {\bibfnamefont {G.}~\bibnamefont {Dubitsky}}, \bibinfo {author} {\bibfnamefont {N.}~\bibnamefont {{Serebryanaya}}}, \bibinfo {author} {\bibfnamefont {M.}~\bibnamefont {Popov}},\ and\ \bibinfo {author} {\bibfnamefont {B.}~\bibnamefont {Sundqvist}},\ }\href {https://doi.org/https://doi.org/10.1016/S0008-6223(97)00234-0} {\bibfield  {journal} {\bibinfo  {journal} {Carbon}\ }\textbf {\bibinfo {volume} {36}},\ \bibinfo {pages} {319} (\bibinfo {year} {1998})}\BibitemShut {NoStop}%
\bibitem [{\citenamefont {Laranjeira}\ \emph {et~al.}(2022)\citenamefont {Laranjeira}, \citenamefont {Marques}, \citenamefont {Melle-Franco}, \citenamefont {Strutyński},\ and\ \citenamefont {Barroso}}]{LARANJEIRA2022}%
  \BibitemOpen
  \bibfield  {author} {\bibinfo {author} {\bibfnamefont {J.}~\bibnamefont {Laranjeira}}, \bibinfo {author} {\bibfnamefont {L.}~\bibnamefont {Marques}}, \bibinfo {author} {\bibfnamefont {M.}~\bibnamefont {Melle-Franco}}, \bibinfo {author} {\bibfnamefont {K.}~\bibnamefont {Strutyński}},\ and\ \bibinfo {author} {\bibfnamefont {M.}~\bibnamefont {Barroso}},\ }\href {https://doi.org/https://doi.org/10.1016/j.carbon.2022.03.055} {\bibfield  {journal} {\bibinfo  {journal} {Carbon}\ }\textbf {\bibinfo {volume} {194}},\ \bibinfo {pages} {297} (\bibinfo {year} {2022})}\BibitemShut {NoStop}%
\bibitem [{\citenamefont {Brazhkin}\ \emph {et~al.}(1997)\citenamefont {Brazhkin}, \citenamefont {Lyapin}, \citenamefont {Popova}, \citenamefont {Voloshin}, \citenamefont {Antonov}, \citenamefont {Lyapin}, \citenamefont {Kluev}, \citenamefont {Naletov},\ and\ \citenamefont {Mel'nik}}]{Braz_clat}%
  \BibitemOpen
  \bibfield  {author} {\bibinfo {author} {\bibfnamefont {V.}~\bibnamefont {Brazhkin}}, \bibinfo {author} {\bibfnamefont {A.}~\bibnamefont {Lyapin}}, \bibinfo {author} {\bibfnamefont {S.}~\bibnamefont {Popova}}, \bibinfo {author} {\bibfnamefont {R.}~\bibnamefont {Voloshin}}, \bibinfo {author} {\bibfnamefont {Y.}~\bibnamefont {Antonov}}, \bibinfo {author} {\bibfnamefont {S.}~\bibnamefont {Lyapin}}, \bibinfo {author} {\bibfnamefont {Y.}~\bibnamefont {Kluev}}, \bibinfo {author} {\bibfnamefont {A.}~\bibnamefont {Naletov}},\ and\ \bibinfo {author} {\bibfnamefont {N.}~\bibnamefont {Mel'nik}},\ }\href {https://doi.org/10.1103/PhysRevB.56.11465} {\bibfield  {journal} {\bibinfo  {journal} {Phys. Rev. B}\ }\textbf {\bibinfo {volume} {56}},\ \bibinfo {pages} {11465} (\bibinfo {year} {1997})}\BibitemShut {NoStop}%
\bibitem [{\citenamefont {Laranjeira}\ and\ \citenamefont {Marques}(2020)}]{LARANJEIRA2020}%
  \BibitemOpen
  \bibfield  {author} {\bibinfo {author} {\bibfnamefont {J.}~\bibnamefont {Laranjeira}}\ and\ \bibinfo {author} {\bibfnamefont {L.}~\bibnamefont {Marques}},\ }\href {https://doi.org/https://doi.org/10.1016/j.mtcomm.2020.100906} {\bibfield  {journal} {\bibinfo  {journal} {Mater. Today Commun.}\ }\textbf {\bibinfo {volume} {23}},\ \bibinfo {pages} {100906} (\bibinfo {year} {2020})}\BibitemShut {NoStop}%
\bibitem [{\citenamefont {Barone}\ \emph {et~al.}(2011)\citenamefont {Barone}, \citenamefont {Hod}, \citenamefont {Peralta},\ and\ \citenamefont {Scuseria}}]{hse}%
  \BibitemOpen
  \bibfield  {author} {\bibinfo {author} {\bibfnamefont {V.}~\bibnamefont {Barone}}, \bibinfo {author} {\bibfnamefont {O.}~\bibnamefont {Hod}}, \bibinfo {author} {\bibfnamefont {J.}~\bibnamefont {Peralta}},\ and\ \bibinfo {author} {\bibfnamefont {G.}~\bibnamefont {Scuseria}},\ }\href {https://doi.org/10.1021/ar100137c} {\bibfield  {journal} {\bibinfo  {journal} {Acc. Chem. Res.}\ }\textbf {\bibinfo {volume} {44}},\ \bibinfo {pages} {269} (\bibinfo {year} {2011})}\BibitemShut {NoStop}%
\bibitem [{\citenamefont {Jhi}\ \emph {et~al.}(2005)\citenamefont {Jhi}, \citenamefont {Louie},\ and\ \citenamefont {Cohen}}]{reentrant_prl_nanotubos}%
  \BibitemOpen
  \bibfield  {author} {\bibinfo {author} {\bibfnamefont {S.-H.}\ \bibnamefont {Jhi}}, \bibinfo {author} {\bibfnamefont {S.}~\bibnamefont {Louie}},\ and\ \bibinfo {author} {\bibfnamefont {M.}~\bibnamefont {Cohen}},\ }\href {https://doi.org/10.1103/PhysRevLett.95.226403} {\bibfield  {journal} {\bibinfo  {journal} {Phys. Rev. Lett.}\ }\textbf {\bibinfo {volume} {95}},\ \bibinfo {pages} {226403} (\bibinfo {year} {2005})}\BibitemShut {NoStop}%
\bibitem [{\citenamefont {Perdew}\ \emph {et~al.}(1996)\citenamefont {Perdew}, \citenamefont {Burke},\ and\ \citenamefont {Ernzerhof}}]{i2}%
  \BibitemOpen
  \bibfield  {author} {\bibinfo {author} {\bibfnamefont {J.}~\bibnamefont {Perdew}}, \bibinfo {author} {\bibfnamefont {K.}~\bibnamefont {Burke}},\ and\ \bibinfo {author} {\bibfnamefont {M.}~\bibnamefont {Ernzerhof}},\ }\href {https://doi.org/10.1103/PhysRevLett.77.3865} {\bibfield  {journal} {\bibinfo  {journal} {Phys. Rev. Lett.}\ }\textbf {\bibinfo {volume} {77}},\ \bibinfo {pages} {3865} (\bibinfo {year} {1996})}\BibitemShut {NoStop}%
\bibitem [{\citenamefont {Perdew}\ \emph {et~al.}(1997)\citenamefont {Perdew}, \citenamefont {Burke},\ and\ \citenamefont {Ernzerhof}}]{i3}%
  \BibitemOpen
  \bibfield  {author} {\bibinfo {author} {\bibfnamefont {J.}~\bibnamefont {Perdew}}, \bibinfo {author} {\bibfnamefont {K.}~\bibnamefont {Burke}},\ and\ \bibinfo {author} {\bibfnamefont {M.}~\bibnamefont {Ernzerhof}},\ }\href {https://doi.org/10.1103/PhysRevLett.78.1396} {\bibfield  {journal} {\bibinfo  {journal} {Phys. Rev. Lett.}\ }\textbf {\bibinfo {volume} {78}},\ \bibinfo {pages} {1396} (\bibinfo {year} {1997})}\BibitemShut {NoStop}%
\bibitem [{\citenamefont {Grimme}\ \emph {et~al.}(2011)\citenamefont {Grimme}, \citenamefont {Ehrlich},\ and\ \citenamefont {Goerigk}}]{grimmed3}%
  \BibitemOpen
  \bibfield  {author} {\bibinfo {author} {\bibfnamefont {S.}~\bibnamefont {Grimme}}, \bibinfo {author} {\bibfnamefont {S.}~\bibnamefont {Ehrlich}},\ and\ \bibinfo {author} {\bibfnamefont {L.}~\bibnamefont {Goerigk}},\ }\href {https://doi.org/https://doi.org/10.1002/jcc.21759} {\bibfield  {journal} {\bibinfo  {journal} {J. Comput. Chem.}\ }\textbf {\bibinfo {volume} {32}},\ \bibinfo {pages} {1456} (\bibinfo {year} {2011})}\BibitemShut {NoStop}%
\bibitem [{\citenamefont {Dovesi}\ \emph {et~al.}(2018)\citenamefont {Dovesi}, \citenamefont {Erba}, \citenamefont {Orlando}, \citenamefont {Zicovich-Wilson}, \citenamefont {Civalleri}, \citenamefont {Maschio}, \citenamefont {Rérat}, \citenamefont {Casassa}, \citenamefont {Baima}, \citenamefont {Salustro},\ and\ \citenamefont {Kirtman}}]{crystal17}%
  \BibitemOpen
  \bibfield  {author} {\bibinfo {author} {\bibfnamefont {R.}~\bibnamefont {Dovesi}}, \bibinfo {author} {\bibfnamefont {A.}~\bibnamefont {Erba}}, \bibinfo {author} {\bibfnamefont {R.}~\bibnamefont {Orlando}}, \bibinfo {author} {\bibfnamefont {C.~M.}\ \bibnamefont {Zicovich-Wilson}}, \bibinfo {author} {\bibfnamefont {B.}~\bibnamefont {Civalleri}}, \bibinfo {author} {\bibfnamefont {L.}~\bibnamefont {Maschio}}, \bibinfo {author} {\bibfnamefont {M.}~\bibnamefont {Rérat}}, \bibinfo {author} {\bibfnamefont {S.}~\bibnamefont {Casassa}}, \bibinfo {author} {\bibfnamefont {J.}~\bibnamefont {Baima}}, \bibinfo {author} {\bibfnamefont {S.}~\bibnamefont {Salustro}},\ and\ \bibinfo {author} {\bibfnamefont {B.}~\bibnamefont {Kirtman}},\ }\href {https://doi.org/https://doi.org/10.1002/wcms.1360} {\bibfield  {journal} {\bibinfo  {journal} {WIREs Comput. Mol. Sci.}\ }\textbf {\bibinfo {volume} {8}},\ \bibinfo {pages} {e1360} (\bibinfo {year} {2018})}\BibitemShut {NoStop}%
\bibitem [{\citenamefont {Sohlberg}\ and\ \citenamefont {Foster}(2020)}]{c60gap}%
  \BibitemOpen
  \bibfield  {author} {\bibinfo {author} {\bibfnamefont {K.}~\bibnamefont {Sohlberg}}\ and\ \bibinfo {author} {\bibfnamefont {M.}~\bibnamefont {Foster}},\ }\href {https://doi.org/10.1039/D0RA07496A} {\bibfield  {journal} {\bibinfo  {journal} {RSC Adv.}\ }\textbf {\bibinfo {volume} {10}},\ \bibinfo {pages} {36887} (\bibinfo {year} {2020})}\BibitemShut {NoStop}%
\bibitem [{\citenamefont {Borlido}\ \emph {et~al.}(2020)\citenamefont {Borlido}, \citenamefont {Schmidt}, \citenamefont {Huran}, \citenamefont {Tran}, \citenamefont {Marques},\ and\ \citenamefont {Botti}}]{mmarques_hsebetterforsolidswithvasp}%
  \BibitemOpen
  \bibfield  {author} {\bibinfo {author} {\bibfnamefont {P.}~\bibnamefont {Borlido}}, \bibinfo {author} {\bibfnamefont {J.}~\bibnamefont {Schmidt}}, \bibinfo {author} {\bibfnamefont {A.}~\bibnamefont {Huran}}, \bibinfo {author} {\bibfnamefont {F.}~\bibnamefont {Tran}}, \bibinfo {author} {\bibfnamefont {M.}~\bibnamefont {Marques}},\ and\ \bibinfo {author} {\bibfnamefont {S.}~\bibnamefont {Botti}},\ }\href {https://doi.org/10.1038/s41524-020-00360-0} {\bibfield  {journal} {\bibinfo  {journal} {Npj Comput. Mater.}\ }\textbf {\bibinfo {volume} {6}},\ \bibinfo {pages} {96} (\bibinfo {year} {2020})}\BibitemShut {NoStop}%
\bibitem [{\citenamefont {Monkhorst}\ and\ \citenamefont {Pack}(1976)}]{monk_grid}%
  \BibitemOpen
  \bibfield  {author} {\bibinfo {author} {\bibfnamefont {H.}~\bibnamefont {Monkhorst}}\ and\ \bibinfo {author} {\bibfnamefont {J.}~\bibnamefont {Pack}},\ }\href {https://doi.org/10.1103/PhysRevB.13.5188} {\bibfield  {journal} {\bibinfo  {journal} {Phys. Rev. B}\ }\textbf {\bibinfo {volume} {13}},\ \bibinfo {pages} {5188} (\bibinfo {year} {1976})}\BibitemShut {NoStop}%
\bibitem [{\citenamefont {Setyawan}\ and\ \citenamefont {Curtarolo}(2010)}]{aflow}%
  \BibitemOpen
  \bibfield  {author} {\bibinfo {author} {\bibfnamefont {W.}~\bibnamefont {Setyawan}}\ and\ \bibinfo {author} {\bibfnamefont {S.}~\bibnamefont {Curtarolo}},\ }\href {https://doi.org/https://doi.org/10.1016/j.commatsci.2010.05.010} {\bibfield  {journal} {\bibinfo  {journal} {Comput. Mater. Sci.}\ }\textbf {\bibinfo {volume} {49}},\ \bibinfo {pages} {299} (\bibinfo {year} {2010})}\BibitemShut {NoStop}%
\bibitem [{\citenamefont {Fleming}\ \emph {et~al.}(1990)\citenamefont {Fleming}, \citenamefont {Siegrist}, \citenamefont {Marsh}, \citenamefont {Hessen}, \citenamefont {Kortan}, \citenamefont {Murphy}, \citenamefont {Haddon}, \citenamefont {Tycko}, \citenamefont {Dabbagh}, \citenamefont {Mujsce}, \citenamefont {Kaplan},\ and\ \citenamefont {Zahurak}}]{fleming}%
  \BibitemOpen
  \bibfield  {author} {\bibinfo {author} {\bibfnamefont {R.}~\bibnamefont {Fleming}}, \bibinfo {author} {\bibfnamefont {T.}~\bibnamefont {Siegrist}}, \bibinfo {author} {\bibfnamefont {P.}~\bibnamefont {Marsh}}, \bibinfo {author} {\bibfnamefont {B.}~\bibnamefont {Hessen}}, \bibinfo {author} {\bibfnamefont {A.}~\bibnamefont {Kortan}}, \bibinfo {author} {\bibfnamefont {D.}~\bibnamefont {Murphy}}, \bibinfo {author} {\bibfnamefont {R.}~\bibnamefont {Haddon}}, \bibinfo {author} {\bibfnamefont {R.}~\bibnamefont {Tycko}}, \bibinfo {author} {\bibfnamefont {G.}~\bibnamefont {Dabbagh}}, \bibinfo {author} {\bibfnamefont {A.}~\bibnamefont {Mujsce}}, \bibinfo {author} {\bibfnamefont {M.}~\bibnamefont {Kaplan}},\ and\ \bibinfo {author} {\bibfnamefont {S.}~\bibnamefont {Zahurak}},\ }\bibfield  {journal} {\bibinfo  {journal} {MRS Proceedings}\ }\textbf {\bibinfo {volume} {206}},\ \href {https://doi.org/10.1557/PROC-206-691} {10.1557/PROC-206-691} (\bibinfo {year} {1990})\BibitemShut {NoStop}%
\bibitem [{\citenamefont {Laranjeira}\ \emph {et~al.}(2018)\citenamefont {Laranjeira}, \citenamefont {Marques}, \citenamefont {Fortunato}, \citenamefont {Melle-Franco}, \citenamefont {Strutyński},\ and\ \citenamefont {Barroso}}]{laran2018}%
  \BibitemOpen
  \bibfield  {author} {\bibinfo {author} {\bibfnamefont {J.}~\bibnamefont {Laranjeira}}, \bibinfo {author} {\bibfnamefont {L.}~\bibnamefont {Marques}}, \bibinfo {author} {\bibfnamefont {N.}~\bibnamefont {Fortunato}}, \bibinfo {author} {\bibfnamefont {M.}~\bibnamefont {Melle-Franco}}, \bibinfo {author} {\bibfnamefont {K.}~\bibnamefont {Strutyński}},\ and\ \bibinfo {author} {\bibfnamefont {M.}~\bibnamefont {Barroso}},\ }\href {https://doi.org/https://doi.org/10.1016/j.carbon.2018.05.070} {\bibfield  {journal} {\bibinfo  {journal} {Carbon}\ }\textbf {\bibinfo {volume} {137}},\ \bibinfo {pages} {511} (\bibinfo {year} {2018})}\BibitemShut {NoStop}%
\bibitem [{\citenamefont {Buga}\ \emph {et~al.}(2000)\citenamefont {Buga}, \citenamefont {Blank}, \citenamefont {Dubitsky}, \citenamefont {Edman}, \citenamefont {Zhu}, \citenamefont {Nyeanchi},\ and\ \citenamefont {Sundqvist}}]{BUGA_1}%
  \BibitemOpen
  \bibfield  {author} {\bibinfo {author} {\bibfnamefont {S.}~\bibnamefont {Buga}}, \bibinfo {author} {\bibfnamefont {V.}~\bibnamefont {Blank}}, \bibinfo {author} {\bibfnamefont {G.}~\bibnamefont {Dubitsky}}, \bibinfo {author} {\bibfnamefont {L.}~\bibnamefont {Edman}}, \bibinfo {author} {\bibfnamefont {X.-M.}\ \bibnamefont {Zhu}}, \bibinfo {author} {\bibfnamefont {E.}~\bibnamefont {Nyeanchi}},\ and\ \bibinfo {author} {\bibfnamefont {B.}~\bibnamefont {Sundqvist}},\ }\href {https://doi.org/https://doi.org/10.1016/S0022-3697(99)00356-X} {\bibfield  {journal} {\bibinfo  {journal} {J. Phys. Chem. Solids}\ }\textbf {\bibinfo {volume} {61}},\ \bibinfo {pages} {1009} (\bibinfo {year} {2000})}\BibitemShut {NoStop}%
\bibitem [{\citenamefont {Buga}\ \emph {et~al.}(2005)\citenamefont {Buga}, \citenamefont {Blank}, \citenamefont {Serebryanaya}, \citenamefont {Dzwilewski}, \citenamefont {Makarova},\ and\ \citenamefont {Sundqvist}}]{BUGA_2}%
  \BibitemOpen
  \bibfield  {author} {\bibinfo {author} {\bibfnamefont {S.}~\bibnamefont {Buga}}, \bibinfo {author} {\bibfnamefont {V.}~\bibnamefont {Blank}}, \bibinfo {author} {\bibfnamefont {N.}~\bibnamefont {Serebryanaya}}, \bibinfo {author} {\bibfnamefont {A.}~\bibnamefont {Dzwilewski}}, \bibinfo {author} {\bibfnamefont {T.}~\bibnamefont {Makarova}},\ and\ \bibinfo {author} {\bibfnamefont {B.}~\bibnamefont {Sundqvist}},\ }\href {https://doi.org/https://doi.org/10.1016/j.diamond.2005.01.018} {\bibfield  {journal} {\bibinfo  {journal} {Diamond and Related Materials}\ }\textbf {\bibinfo {volume} {14}},\ \bibinfo {pages} {896} (\bibinfo {year} {2005})},\ \bibinfo {note} {proceedings of Diamond 2004, the 15th European Conference on Diamond, Diamond-Like Materials, Carbon Nanotubes, Nitrides and Silicon Carbide}\BibitemShut {NoStop}%
\bibitem [{\citenamefont {Zipoli}\ and\ \citenamefont {Bernasconi}(2008)}]{Bernasconi}%
  \BibitemOpen
  \bibfield  {author} {\bibinfo {author} {\bibfnamefont {F.}~\bibnamefont {Zipoli}}\ and\ \bibinfo {author} {\bibfnamefont {M.}~\bibnamefont {Bernasconi}},\ }\href {https://doi.org/10.1103/PhysRevB.77.115432} {\bibfield  {journal} {\bibinfo  {journal} {Phys. Rev. B}\ }\textbf {\bibinfo {volume} {77}},\ \bibinfo {pages} {115432} (\bibinfo {year} {2008})}\BibitemShut {NoStop}%
\bibitem [{\citenamefont {Yang}\ \emph {et~al.}(2007)\citenamefont {Yang}, \citenamefont {Tse},\ and\ \citenamefont {Iitaka}}]{tse}%
  \BibitemOpen
  \bibfield  {author} {\bibinfo {author} {\bibfnamefont {J.}~\bibnamefont {Yang}}, \bibinfo {author} {\bibfnamefont {J.}~\bibnamefont {Tse}},\ and\ \bibinfo {author} {\bibfnamefont {T.}~\bibnamefont {Iitaka}},\ }\href {https://doi.org/10.1063/1.2771162} {\bibfield  {journal} {\bibinfo  {journal} {J. Chem. Phys.}\ }\textbf {\bibinfo {volume} {127}},\ \bibinfo {pages} {134906} (\bibinfo {year} {2007})}\BibitemShut {NoStop}%
\bibitem [{\citenamefont {Laranjeira}\ \emph {et~al.}(2023)\citenamefont {Laranjeira}, \citenamefont {Errea}, \citenamefont {Dangić}, \citenamefont {Marques}, \citenamefont {Melle-Franco},\ and\ \citenamefont {Strutyński}}]{Laranjeira2023}%
  \BibitemOpen
  \bibfield  {author} {\bibinfo {author} {\bibfnamefont {J.}~\bibnamefont {Laranjeira}}, \bibinfo {author} {\bibfnamefont {I.}~\bibnamefont {Errea}}, \bibinfo {author} {\bibfnamefont {D.}~\bibnamefont {Dangić}}, \bibinfo {author} {\bibfnamefont {L.}~\bibnamefont {Marques}}, \bibinfo {author} {\bibfnamefont {M.}~\bibnamefont {Melle-Franco}},\ and\ \bibinfo {author} {\bibfnamefont {K.}~\bibnamefont {Strutyński}},\ }\href {https://doi.org/10.1002/pssr.202300249} {\bibfield  {journal} {\bibinfo  {journal} {Phys. status solidi -RRL}\ ,\ \bibinfo {pages} {2300249}} (\bibinfo {year} {2023})}\BibitemShut {NoStop}%
\bibitem [{\citenamefont {Burgos}\ \emph {et~al.}(2000)\citenamefont {Burgos}, \citenamefont {Halac}, \citenamefont {Weht}, \citenamefont {Bonadeo}, \citenamefont {Artacho},\ and\ \citenamefont {Ordej\'on}}]{Burgos}%
  \BibitemOpen
  \bibfield  {author} {\bibinfo {author} {\bibfnamefont {E.}~\bibnamefont {Burgos}}, \bibinfo {author} {\bibfnamefont {E.}~\bibnamefont {Halac}}, \bibinfo {author} {\bibfnamefont {R.}~\bibnamefont {Weht}}, \bibinfo {author} {\bibfnamefont {H.}~\bibnamefont {Bonadeo}}, \bibinfo {author} {\bibfnamefont {E.}~\bibnamefont {Artacho}},\ and\ \bibinfo {author} {\bibfnamefont {P.}~\bibnamefont {Ordej\'on}},\ }\href {https://doi.org/10.1103/PhysRevLett.85.2328} {\bibfield  {journal} {\bibinfo  {journal} {Phys. Rev. Lett.}\ }\textbf {\bibinfo {volume} {85}},\ \bibinfo {pages} {2328} (\bibinfo {year} {2000})}\BibitemShut {NoStop}%
\end{thebibliography}%
\onecolumngrid
\includegraphics[scale=1]{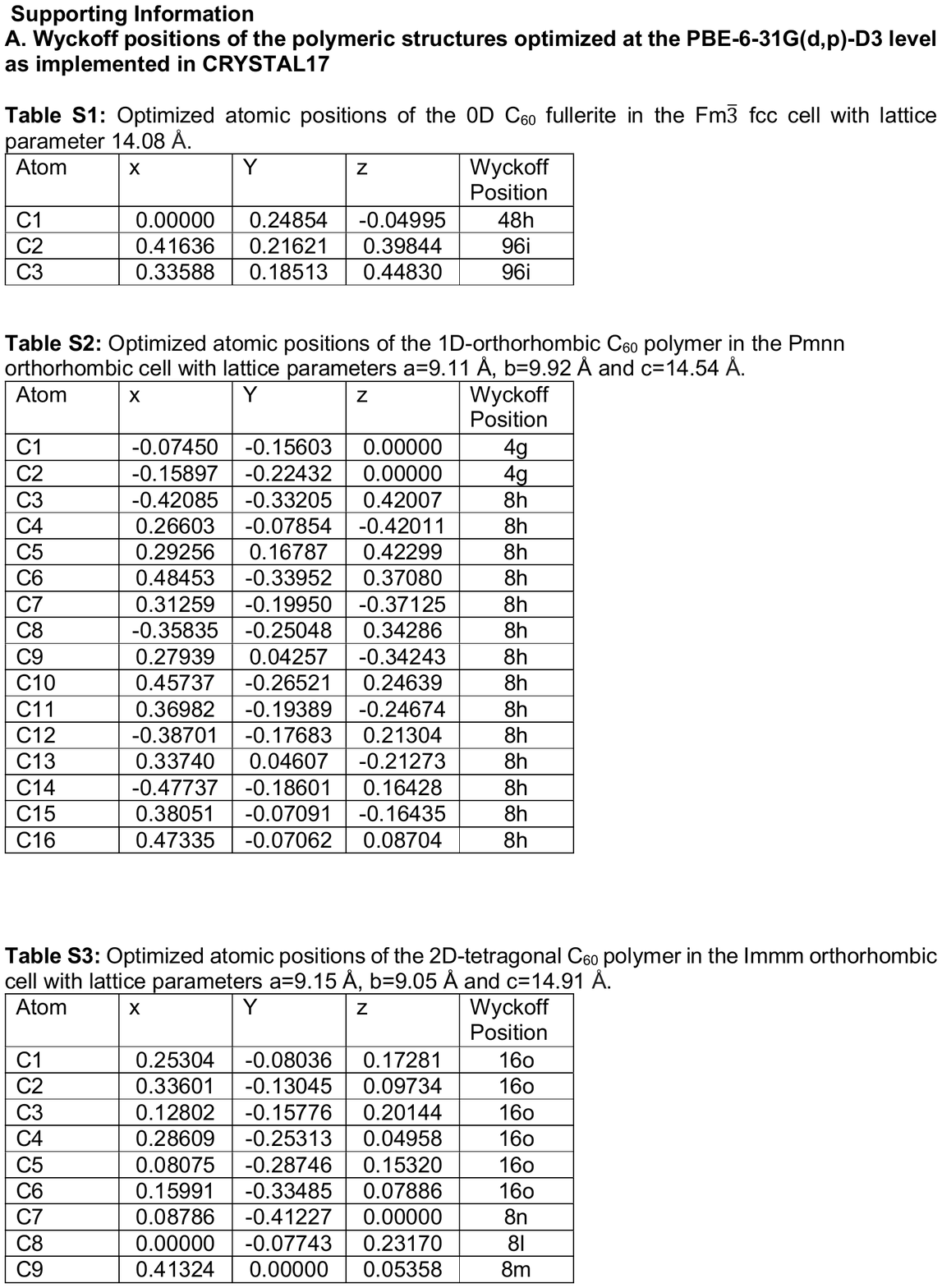}
\includegraphics[scale=1]{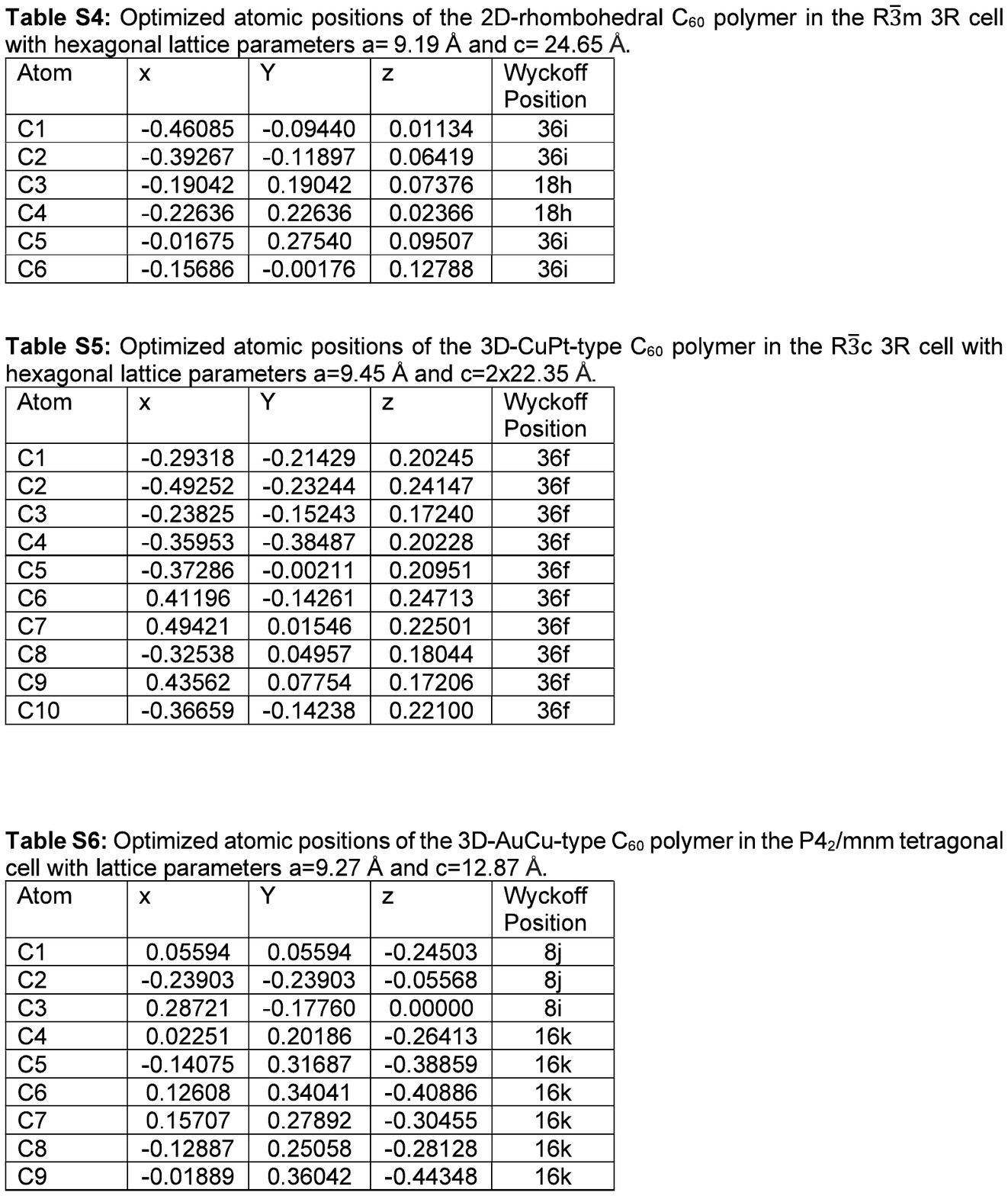}
\includegraphics[scale=1]{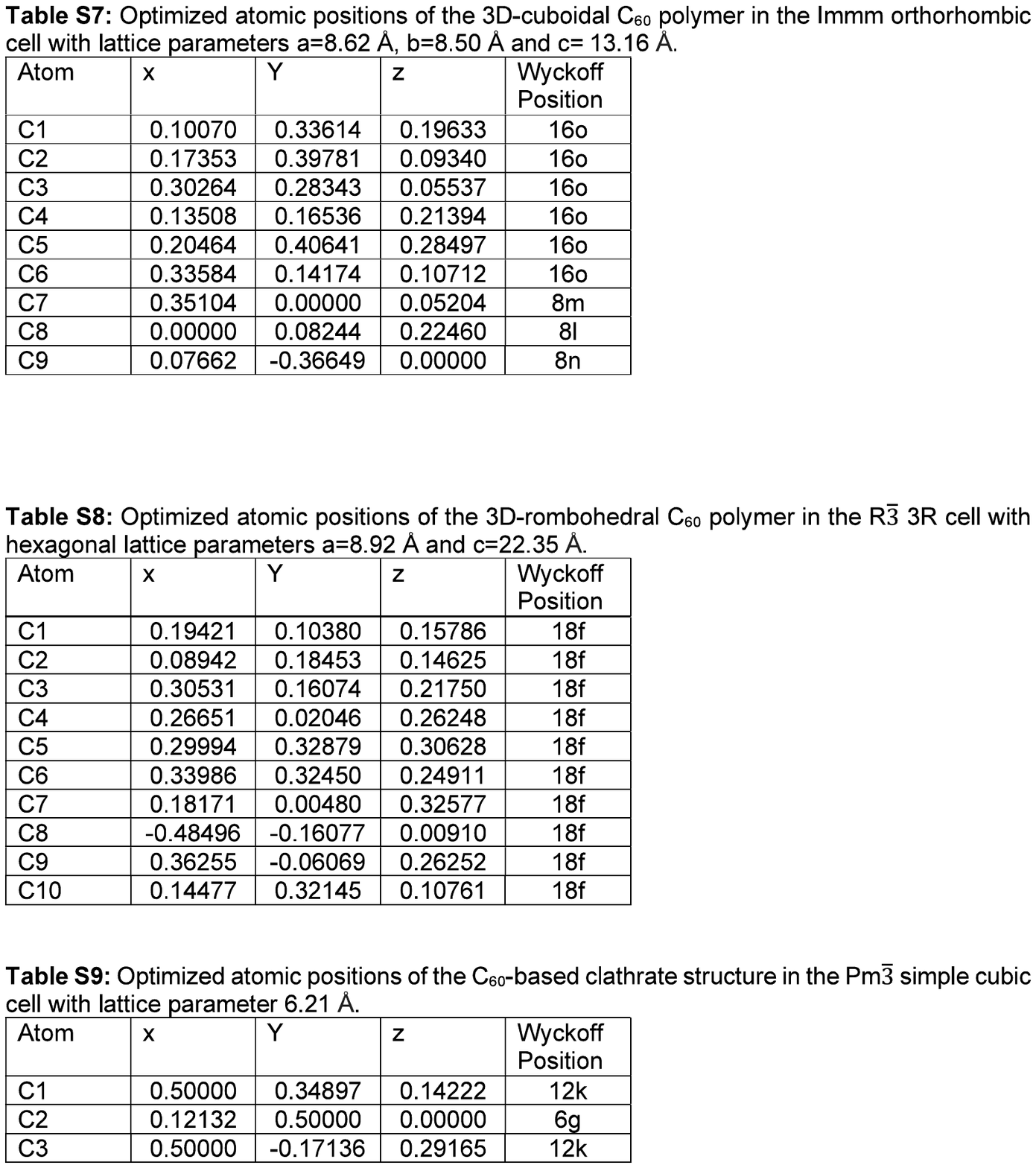}
\includegraphics[scale=1]{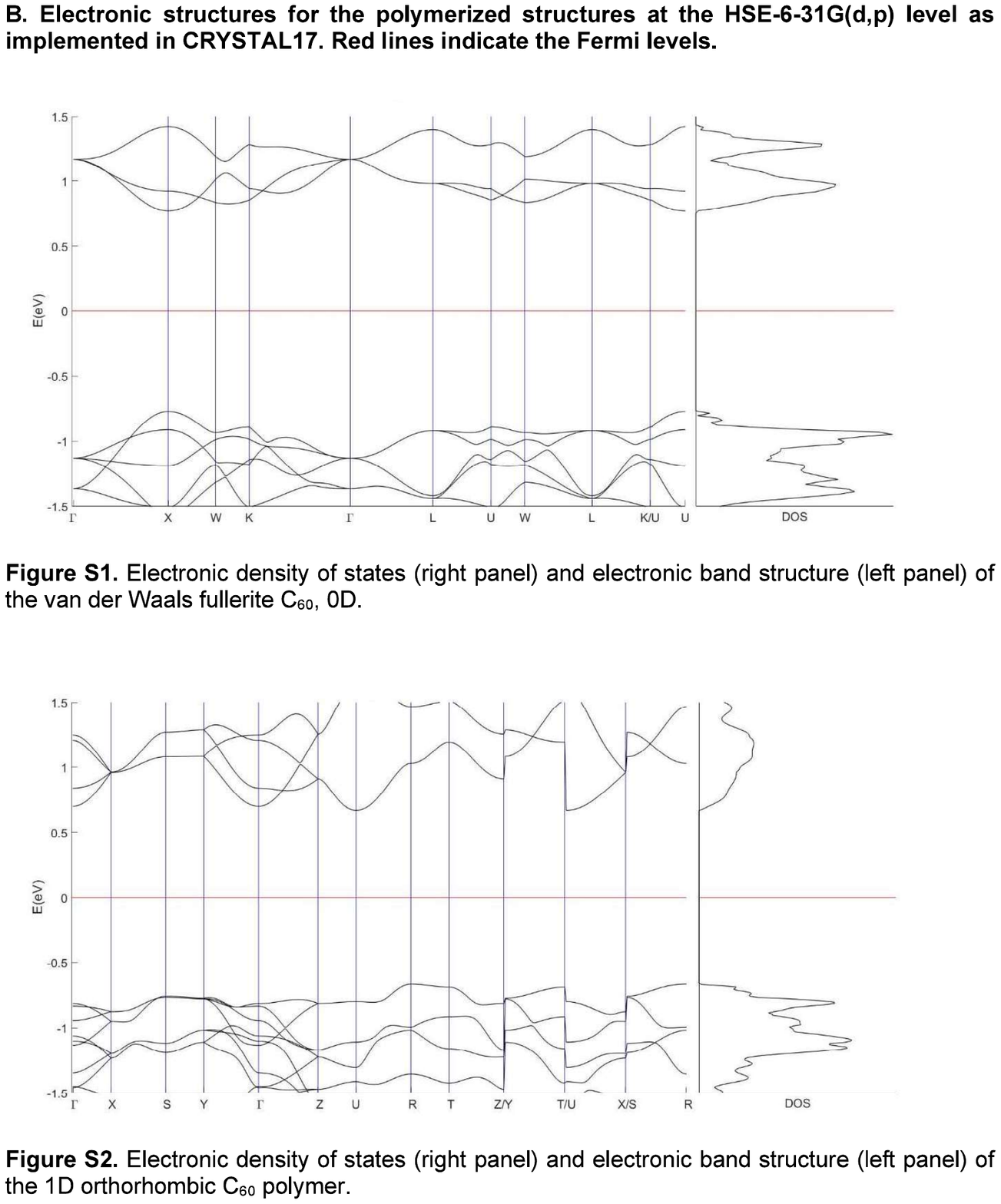}
\includegraphics[scale=1]{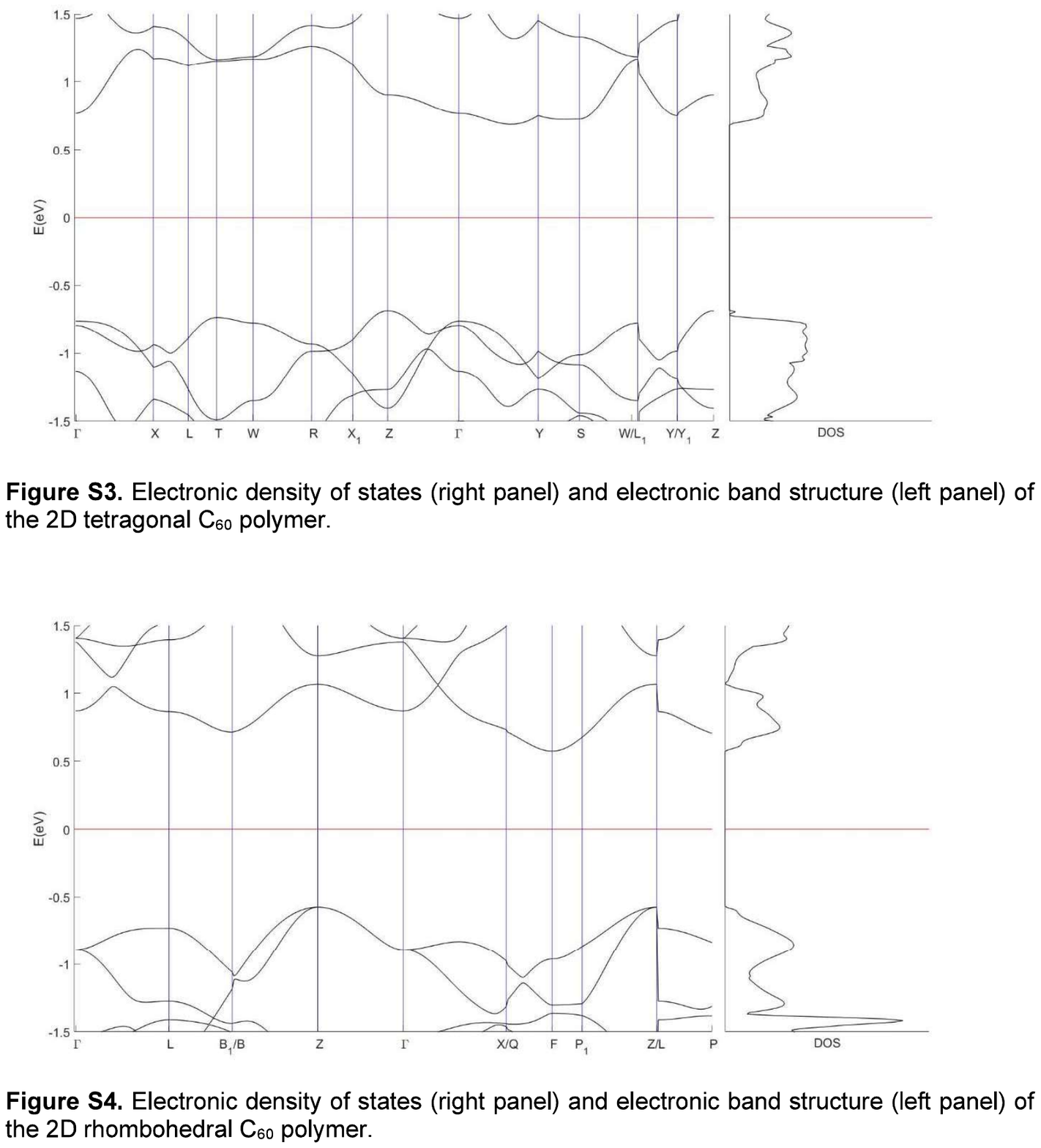}
\includegraphics[scale=1]{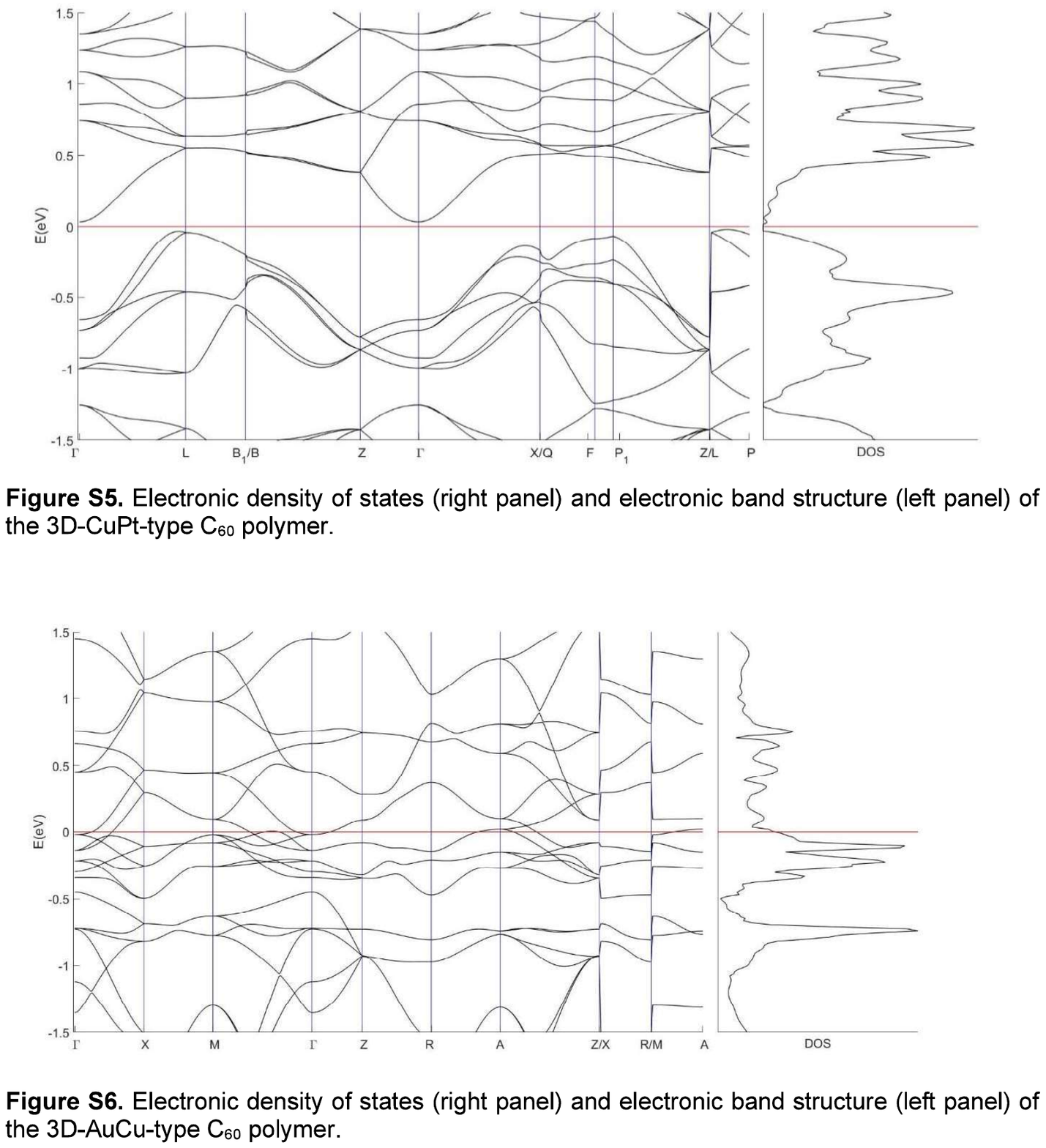}
\includegraphics[scale=1]{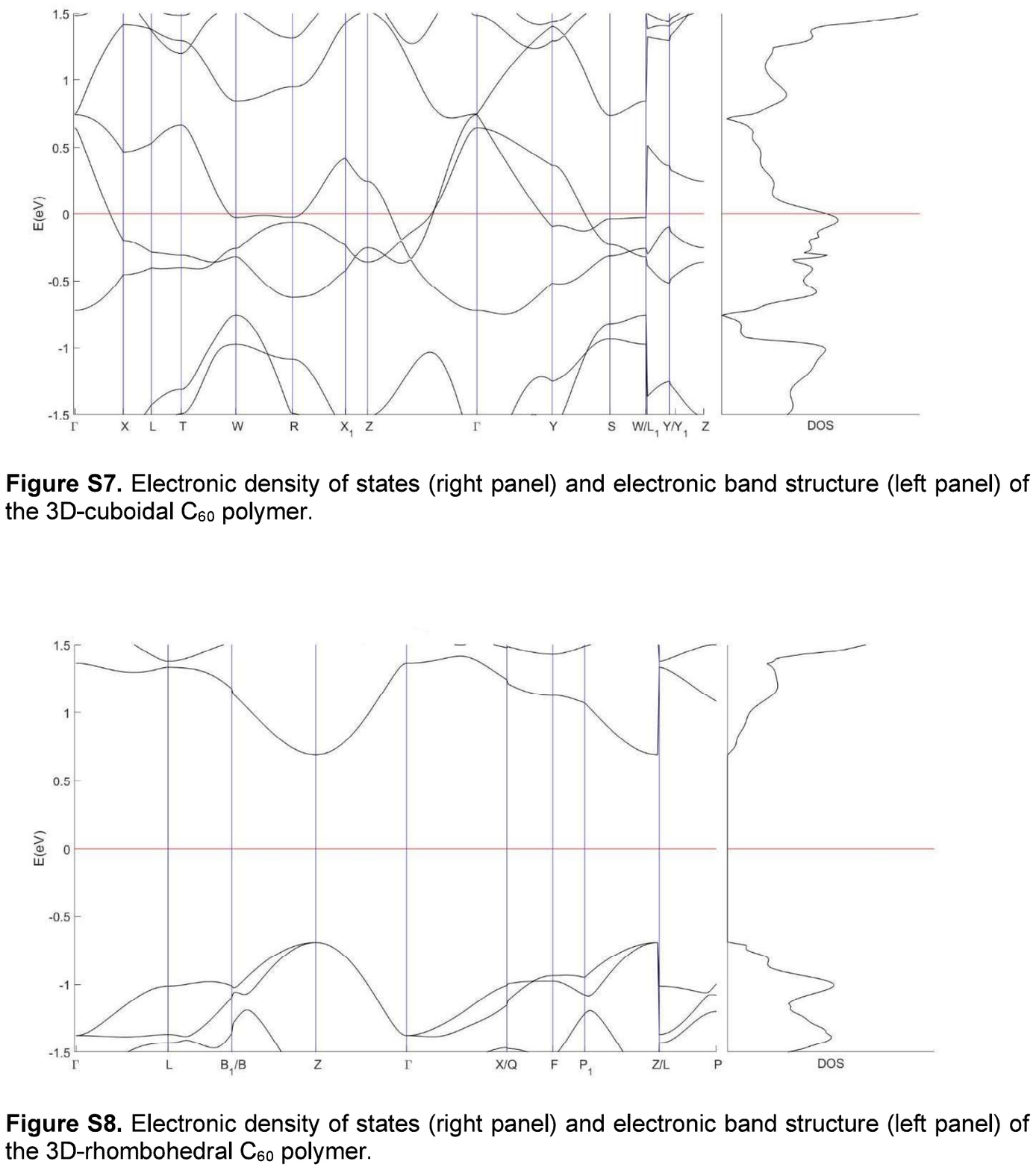}
\includegraphics[scale=1]{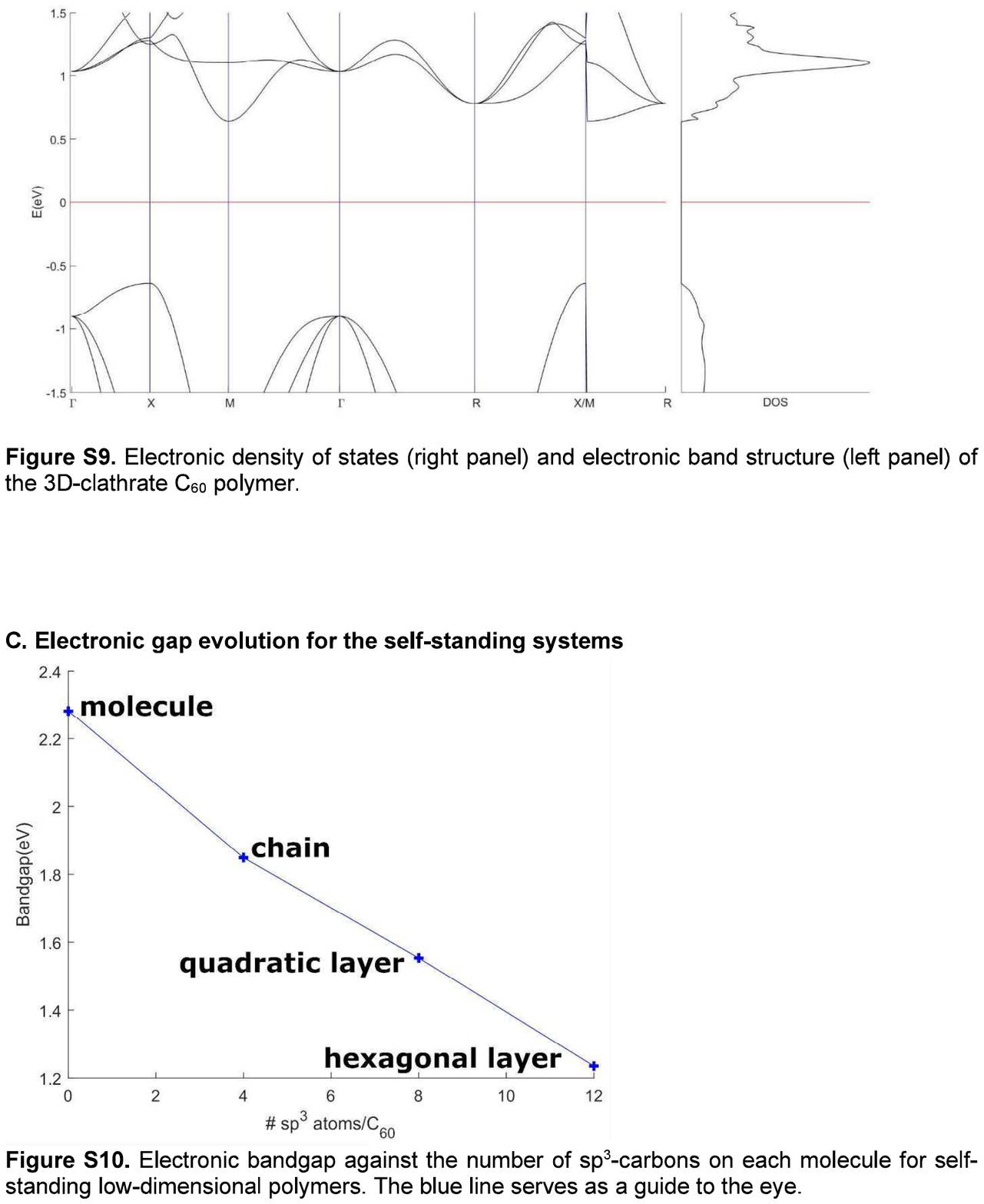}
\includegraphics[scale=1]{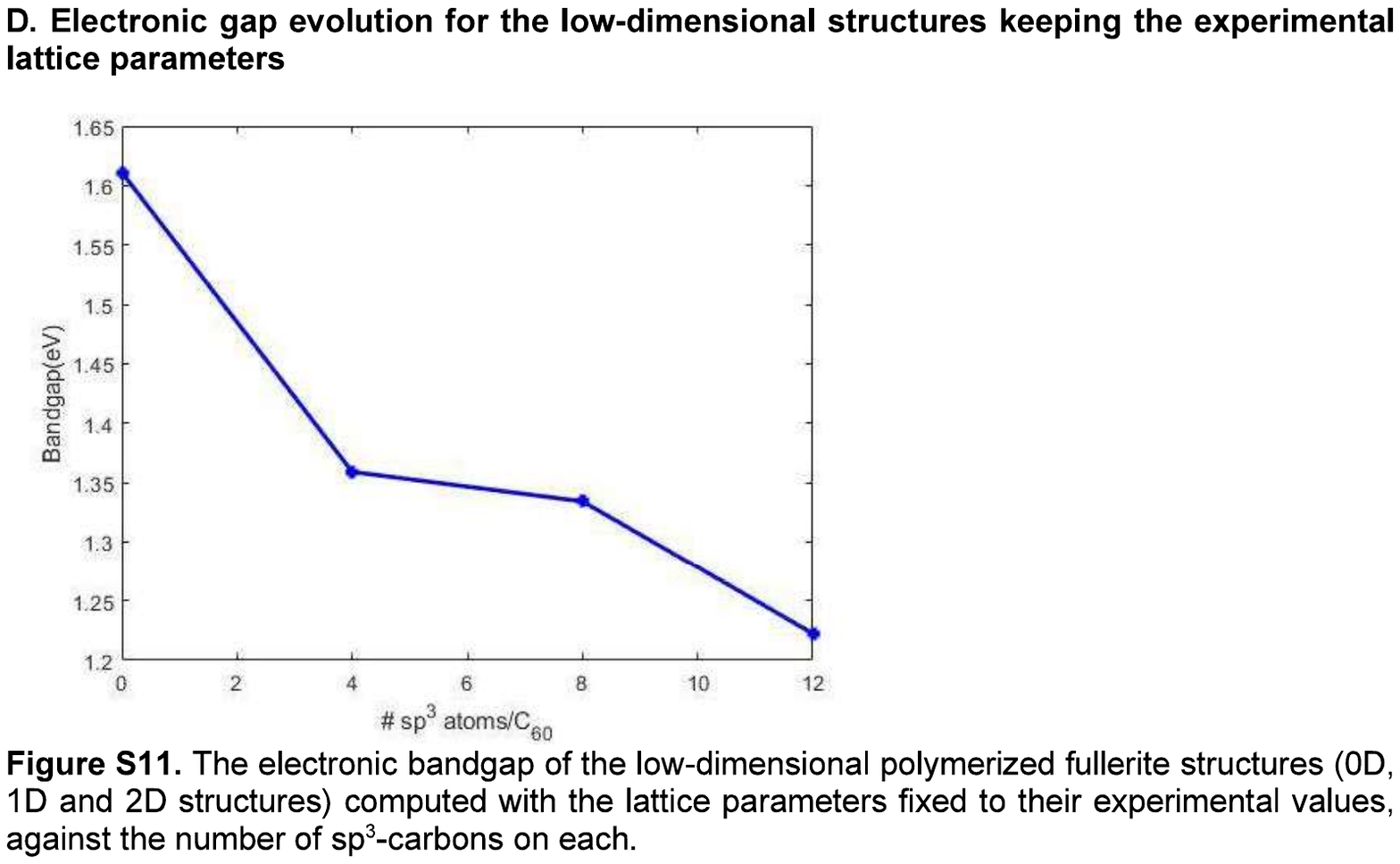}

\end{document}